\documentclass[12pt]{article}

\usepackage{amsmath,amsfonts,amsthm,amssymb}
\usepackage{subfigure}
\usepackage{setspace}
\usepackage{Tabbing}
\usepackage{lastpage}
\usepackage{extramarks}
\usepackage{chngpage}
\usepackage[usenames,dvipsnames]{color}
\usepackage{float,wrapfig}
\usepackage[dvips]{graphicx}
\usepackage[colorlinks,hyperindex]{hyperref}
\hypersetup{citecolor=Magenta, linkcolor=Blue, urlcolor=Red}
 \usepackage[usenames,dvipsnames]{pstricks}
 \usepackage{epsfig}
 \usepackage{pst-grad} 
 \usepackage{pst-plot} 

\def\br{\begin{eqnarray}}
\def\er{\end{eqnarray}}
\def\be{\begin{equation}}
\def\ee{\end{equation}}
\def\({\left(}
\def\){\right)}
\def\rlx{\relax\leavevmode}
\def\IR{\rlx\hbox{\rm I\kern-.18em R}}

	\newcommand{\ba}{\begin{array}}
	\newcommand{\ea}{\end{array}}
	\newcommand{\beqa}{\begin{equation}\begin{array}{rcl}}
	\newcommand{\eeqa}[1]{\end{array}\label{#1}\end{equation}}

\def\IZ{\rlx\hbox{\sf Z\kern-.4em Z}}
\def\IR{\rlx\hbox{\rm I\kern-.18em R}}
\def\IC{\rlx\hbox{\,$\inbar\kern-.3em{\rm C}$}}
\def\one{\hbox{{1}\kern-.25em\hbox{l}}}

\renewcommand{\theequation}{\thesection.\arabic{equation}}
\def\IZ{\rlx\hbox{\sf Z\kern-.4em Z}}
\def\IR{\rlx\hbox{\rm I\kern-.18em R}}
\def\IC{\rlx\hbox{\,$\inbar\kern-.3em{\rm C}$}}
\def\one{\hbox{{1}\kern-.25em\hbox{l}}}
     
\begin{document}
\newcommand{\sect}[1]{\setcounter{equation}{0}\section{#1}}
\renewcommand{\theequation}{\thesection.\arabic{equation}}



     
     
     
     

\newcommand{\m}{ \mu}
\newcommand{\la}{ \lambda}
\newcommand{\C}{ C_0 }

\newcommand{\s}{ \sigma}
\newcommand{\ka}{ \kappa}

\newcommand{\e}{ \varepsilon}

\newcommand{\spa}{$\spadesuit$}
\newcommand{\bul}{$\bullet$}

\newcommand{\ir}{{\mathrm{IR}}}
\newcommand{\uv}{{\mathrm{UV}}}

\def\PRL#1#2#3{{\sl Phys. Rev. Lett.} {\bf#1} (#2) #3}
\def\NPB#1#2#3{{\sl Nucl. Phys.} {\bf B#1} (#2) #3}
\def\NPBFS#1#2#3#4{{\sl Nucl. Phys.} {\bf B#2} [FS#1] (#3) #4}
\def\CMP#1#2#3{{\sl Commun. Math. Phys.} {\bf #1} (#2) #3}
\def\PRD#1#2#3{{\sl Phys. Rev.} {\bf D#1} (#2) #3}
\def\PRB#1#2#3{{\sl Phys. Rev.} {\bf B#1} (#2) #3}

\def\PLA#1#2#3{{\sl Phys. Lett.} {\bf #1A} (#2) #3}
\def\PLB#1#2#3{{\sl Phys. Lett.} {\bf #1B} (#2) #3}
\def\JMP#1#2#3{{\sl J. Math. Phys.} {\bf #1} (#2) #3}
\def\PTP#1#2#3{{\sl Prog. Theor. Phys.} {\bf #1} (#2) #3}
\def\SPTP#1#2#3{{\sl Suppl. Prog. Theor. Phys.} {\bf #1} (#2) #3}
\def\AoP#1#2#3{{\sl Ann. of Phys.} {\bf #1} (#2) #3}
\def\PNAS#1#2#3{{\sl Proc. Natl. Acad. Sci. USA} {\bf #1} (#2) #3}
\def\RMP#1#2#3{{\sl Rev. Mod. Phys.} {\bf #1} (#2) #3}
\def\PR#1#2#3{{\sl Phys. Reports} {\bf #1} (#2) #3}
\def\AoM#1#2#3{{\sl Ann. of Math.} {\bf #1} (#2) #3}
\def\UMN#1#2#3{{\sl Usp. Mat. Nauk} {\bf #1} (#2) #3}
\def\FAP#1#2#3{{\sl Funkt. Anal. Prilozheniya} {\bf #1} (#2) #3}
\def\FAaIA#1#2#3{{\sl Functional Analysis and Its Application} {\bf #1} (#2)
#3}
\def\BAMS#1#2#3{{\sl Bull. Am. Math. Soc.} {\bf #1} (#2) #3}
\def\TAMS#1#2#3{{\sl Trans. Am. Math. Soc.} {\bf #1} (#2) #3}
\def\InvM#1#2#3{{\sl Invent. Math.} {\bf #1} (#2) #3}
\def\LMP#1#2#3{{\sl Letters in Math. Phys.} {\bf #1} (#2) #3}
\def\IJMPA#1#2#3{{\sl Int. J. Mod. Phys.} {\bf A#1} (#2) #3}
\def\AdM#1#2#3{{\sl Advances in Math.} {\bf #1} (#2) #3}
\def\RMaP#1#2#3{{\sl Reports on Math. Phys.} {\bf #1} (#2) #3}
\def\IJM#1#2#3{{\sl Ill. J. Math.} {\bf #1} (#2) #3}
\def\APP#1#2#3{{\sl Acta Phys. Polon.} {\bf #1} (#2) #3}
\def\TMP#1#2#3{{\sl Theor. Mat. Phys.} {\bf #1} (#2) #3}
\def\JPA#1#2#3{{\sl J. Physics} {\bf A#1} (#2) #3}
\def\JSM#1#2#3{{\sl J. Soviet Math.} {\bf #1} (#2) #3}
\def\MPLA#1#2#3{{\sl Mod. Phys. Lett.} {\bf A#1} (#2) #3}
\def\JETP#1#2#3{{\sl Sov. Phys. JETP} {\bf #1} (#2) #3}
\def\JETPL#1#2#3{{\sl  Sov. Phys. JETP Lett.} {\bf #1} (#2) #3}
\def\PHSA#1#2#3{{\sl Physica} {\bf A#1} (#2) #3}
\def\PHSD#1#2#3{{\sl Physica} {\bf D#1} (#2) #3}

\begin{titlepage}
\vspace*{-2 cm}
\noindent
\begin{flushright}
\end{flushright}

\vskip 1 cm
\begin{center}
{\Large\bf  Domain Walls in Extended Lovelock Gravity } \vglue 1  true cm

  {U. Camara dS}$^{*}$\footnote {e-mail: ulyssescamara@gmail.com}, {C.P. Constantinidis}$^{*}$\footnote {e-mail: cpconstantinidis@gmail.com}, { A.L. Alves Lima}$^{*}$\footnote {e-mail: andrealves.fis@gmail.com} and { G.M.Sotkov}$^{*}$\footnote {e-mail: sotkov@cce.ufes.br, gsotkov@yahoo.com.br}\\

\vspace{1 cm}

${}^*\;${\footnotesize Departamento de F\'\i sica - CCE\\
Universidade Federal de Espirito Santo\\
29075-900, Vitoria - ES, Brazil}\\

\vspace{0,5 cm}

\vspace{1 cm}

\end{center}

\normalsize
\vskip 0.5cm

\begin{center}
{ {\bf ABSTRACT}}\\
\end{center}

\vspace{.5cm}

 We derive a BPS-like first order system of equations for a family of flat static domain walls (DWs) of dimensionally extended cubic Lovelock Gravity  coupled to massive scalar self-interacting matter. The explicit construction of such DWs is achieved by introducing of an appropriate matter superpotential. We further analyse the dependence of the geometric properties of the  asymptotically $AdS_d$ space-times representing  distinct  DWs  on  the shape of the matter potential, on the values of the Lovelock couplings and on the scalar field boundary conditions. Few explicit  examples of Lovelock DWs interpolating between $AdS$-type vacua of different cosmological constants are presented. In five dimensions 
 our method provides interesting solutions of the Myers-Robinson Quasi-topological Gravity in the presence of matter important for the description of the specific renormalization group flows in its holographic dual four-dimensional $CFT$ perturbed by relevant operators.

\end{titlepage}

\tableofcontents 

\setcounter{equation}{0}
\section{Introduction}

Recent investigations of the  positivity and causality properties of the energy fluxes in certain  $N=1$ and $N=2$ supersymmetric (and of a family of  non-supersymmetric)  four-dimensional CFT's \cite{hydro}, \cite{maldahof} having two distinct central charges $a\neq c$, make evident that their strong coupling $AdS/CFT$ description requires the knowledge of appropriate classical solutions of the Gauss-Bonnet and dimensionally extended cubic Lovelock $d=5$ (super) Gravity \cite{shydro}, \cite{hidro},\cite{2},\cite{bala}. An important physical condition on the particular form of the actions of such  generalizations of the Einstein Gravity is that the equations  for a large class of solutions, as for example the black holes and domain walls, as well as the ones  for linear fluctuations of the metrics around such backgrounds, to be of \emph{second order} \cite{love},\cite{deser}. As it is well known, this requirement can be (partially) achieved by selecting very specific combinations of the terms quadratic and cubic in $R_{abcd}$, $R_{ab}$ and $R$ of (quasi)topological nature, whose form is strongly dependent on the dimension of the space-time. The  $d > 4$ cubic Quasi-Topological Gravity \cite{My_qtop},\cite{c_th} represents the most physical higher order extended Gravity available in five dimensions. Let us mention few of its characteristic features: 
(a) It admits simple black hole solutions, (b) the linear fluctuations of the metrics around arbitrary background involve at most up to fourth order in the derivatives of the metrics, while the corresponding fluctuations around \emph{conformally flat} solutions are of \emph{second} order, (c)  its black holes solutions  provide dual description of a family of $d=4$ (finite temperature) CFT's having the most general stress tensor 3-point function (with $t_4 \neq 0$ \cite{shydro}, \cite{My_thol}, \cite{2},\cite{bala}, \cite{espanha}).

The present paper is devoted to the problem of the explicit construction of a family of  static flat domain walls solutions (DWs) of the cubic  Quasi-Topological Gravity coupled to massive scalar self-interacting matter in arbitrary $d\ge 3$ dimensions. Similarly to the case of the $AdS$/$CFT$ correspondence \cite{witt} involving only the Einstein Gravity with negative cosmological constant,  such DWs solutions of the extended Lovelock Gravity are expected to describe the strong coupling behavior of  certain zero temperature CFTs as well as the holographic renormalization group (RG) flows  between different $(d-1)$-dimensional CFTs \cite{2},\cite{VVB},\cite{rg}. Our \emph{main result} consists in the introduction of an appropriate matter superpotential that enable us to derive the specific BPS-like first order system of equations for the corresponding flat Lovelock DWs. They appear to be a natural generalization of the superpotential constructions for $d\geq 3$ Gauss-Bonnet Gravity coupled to scalar matter, proposed in ref.\cite{nmg}, following the idea of ref.\cite{zee}. We further analyze the geometric properties of the  asymptotically $AdS_d$ space-times representing all the allowed regular and singular DWs as a function of the shape of the matter potential, of the values of the Lovelock couplings and of the scalar field boundary conditions. Few explicit  examples of Lovelock DWs corresponding to particular simple choices of the superpotential are presented.


\section{Superpotential for Lovelock Domain Walls}

As in the case of Einstein Gravity coupled to scalar matter \cite{5},\cite{6},\cite{town2} all the static flat DWs solutions of the extended cubic Lovelock Gravity \cite{My_qtop} are defined by the anzatz: 
\begin{eqnarray}
ds^2_d&=&g_{\mu\nu}(x_{\rho})dx^{\mu}dx^{\nu}=dy^2+e^{2A(y)}\eta_{ij}dx^idx^j ,\quad\mu,\nu=0,1,2..., d - 1    \nonumber\\ 
\sigma(x_i,y)&=&\sigma(y),\quad \eta_{ij}=\big(-,+,...,+\big),\quad i,j=0,1...,d-2 \label{dw}
\end{eqnarray} 
and satisfying  the following boundary conditions (b.c.):
\begin{eqnarray}
e^{A}(y\rightarrow\pm\infty) \approx e^{\frac{y}{L_{\pm}}},\quad\quad\ \sigma(\pm\infty)=\sigma^*_{\pm} \label{bc}
\end{eqnarray} 
 These b.c's are specific for the case when we have at least two different $AdS_d$-type of vacua ${\Lambda_{\pm}=-(d-1)(d-2)/2L^2_{\pm}}$ and  $\sigma_{\pm}^*$  with $V^{'}(\sigma_{\pm}^*)=0$  \cite{6}. Then the resulting  DWs  represent smooth solutions of the extended Lovelock-matter model interpolating between two such vacua configurations placed at the causal limits, i.e. $y \rightarrow \pm \infty$ ends (boundaries and/or horizons) of the effective (non-constant curvature) asymptotically $AdS_d$ space-times, called $(a)AdS_d$.

\subsection{Extended $d\ge 3$ cubic Lovelock gravity $II^{nd}$ order equations} 

 Let us start by reminding the well known  fact that such DWs represent conformally flat metrics, i.e. their Weyl tensor is vanishing and therefore the corresponding Riemann tensor $R_{\mu\nu\rho\tau}$ can be realised in terms of the Ricci tensor $R_{\mu\nu}$,the scalar curvature R and the metrics $g_{\mu\nu}$ only. As a consequence all the quadratic and cubic terms in the Quasi-topological Gravity action \cite{My_qtop} containing different contractions of one,two or three Riemann tensors can be reduced to following set of five invariants: $R^2$,$R_{\mu\nu}R^{\mu\nu}$ and $R^3$, $RR_{\mu\nu}R^{\mu\nu}$,$R_{\mu\nu}R^{\mu\rho}R_{\rho}^{\mu}$. Similarly to the pure Gauss-Bonnet case \cite{zee},\cite{nmg},\cite{3} we can use as a starting point of our derivation of the DWs equations the following effective (simplified) Lovelock-matter action (for all $d\geq 3$):
\begin{eqnarray}
&&S^{{\mathrm{eff}}}_{GBL}= \frac{1}{\kappa^2}\int \! d^d x \, \sqrt{-g} \, \Bigg[ R + \frac{\lambda_0}{m^2} \left(R^{\mu\nu} R_{\mu \nu} - \gamma_d R^2 \right) + \label{gbl-action}\\
&&+ \frac{d(d+4)-4}{(d-1)^2 (d-2)^2} \frac{\mu_0}{(m^2)^2} \left( R^3 + \alpha_d R R^{\mu\nu} R_{\mu\nu} + \beta_d R^{\mu}_{\nu} R^{\nu}_{\rho} R^{\rho}_{\mu} \right) - \kappa^2\left(\frac{1}{2} g^{\mu\nu}\partial_{\mu} \sigma \, \partial_{\nu} \sigma + V(\sigma)\right) \Bigg] ,\nonumber
\end{eqnarray}
where $\kappa^2=\frac{8\pi^{(d-1)/2}}{\Gamma((d-1)/2)} G_d$ is defining the Plank scale, the $m^2\sim\frac{1}{L^2}$ is representing the new scale (related to the bare cosmological constant\footnote{The negative bare cosmological constant $\Lambda$ is implicitly defined by the vacuum value of the matter potential $\kappa^2 V(\sigma^*)=2\Lambda=-(d-1)(d-2)/L^2$ and should be distinguished from the effective cosmological constant $\Lambda^{{\mathrm{eff}}}$ introduced in Sect.3.}) specific for the ``higher order'' gravitational models \cite{deser}, while  $\lambda_0$ and $\mu_0$ denote the appropriately normalized dimensionless Gauss-Bonnet and Lovelock ``gravitational'' couplings. The free parameters $\gamma_d$, $\alpha_d$ and $\beta_d$ are further determined in App.A by the requirement that the equations derived from the above action (\ref{gbl-action}) when written for flat DWs metrics (\ref{dw}) to become of \emph{second order}:   
\begin{eqnarray}
\gamma_d=\frac{d}{4(d-1)},\quad \alpha_d=-\frac{12d(d-1)}{d(d+4)-4}, \quad \beta_d=\frac{16(d-1)^2}{d(d+4)-4},\label{valores}
\end{eqnarray}
As expected the value of $\gamma_d$ coincides to the one that appears in the effective action for flat DWs of the pure Gauss-Bonnet Gravity and also in the dimensional extension of the New Massive Gravity (NMG) \cite{nmg}), while all the $d=3$ parameters values $\alpha_3 = - \frac{72}{17}$, $\beta_3 = \frac{64}{17}$ are equal the coefficients of the cubic extension of the three dimensional NMG considered by Sinha \cite{sinha}. 

Instead of directly substituting the DW metrics (\ref{dw}) in the equations of motion derived from the action (\ref{gbl-action}), we chose to work with the equivalent (but more efficient and rather economic) effective Lagrangian method\footnote{See for example refs. \cite{town1}, \cite{town2} for the case of DWs in Einstein gravity.} properly adapted to the case of ``higher order derivatives" gravitational models in our App.A. It consists in the introduction of an arbitrary function $f(y)$ (sort of ``lapse") in the definition of the DWs metrics (\ref{dw}):
\begin{equation}
ds^2 = f^2(y) e^{2 (d-1) A(y)} dy^2 + e^{2A(y)} \eta_{ij} dx^i dx^j      \label{met}
\end{equation}
reflecting the freedom we have in choosing the DWs ``radial" coordinate $y$. As it is shown in App.A., by  substituting this ``modified" DWs ansatz (\ref{met}) in the action (\ref{gbl-action}) and next by imposing  the condition for eliminating  all the higher derivatives terms we find  the particular values (\ref{valores}) of the parameters $\gamma_d$, $\alpha_d$ and $\beta_d$. As a result we derive the following action $\kappa^2S = \int \! d^{d-1}x \,  dy \; {\mathcal{L}}^{{\mathrm{eff}}}$, where  the ``effective" DWs Lagrangian (modulo total derivatives) is given by:
\begin{eqnarray}
&{\mathcal{L}}^{\mathrm{eff}} = \frac{1}{2 f(y)} \left[ 2(d-1)(d-2) \dot{A}^2(y) - \kappa^2 \dot{\sigma}^2(y) \right] -  f(y) \, e^{2(d-1)A(y)} \kappa^2 V(\sigma)   \nonumber\\
&+ \lambda_0 \frac{(d-1)(d-2)^2(d-4)}{12}\frac{e^{-2(d-1)A(y)}}{ m^2f^3(y)} \dot{A}^4(y) + \mu_0 \frac{(d-1)(d-2)^2(d-6)}{5m^4} \frac{e^{-4(d-1)A(y)}}{f^5(y)}  \dot{A}^6       \label{L-eff}
\end{eqnarray}
The above ``mechanical" system is indeed constrained since ${\mathcal{L}}^{{\mathrm{eff}}}$ does not depend on the $f(y)$ derivatives and as a consequence the variation with respect to $f(y)$  gives the constraint (\ref{constr}). The corresponding second order equations describing the Lovelock DWs (derived in App.A from the Lagrangian (\ref{L-eff})) have  the following  compact and rather simple equivalent (to be compared with eqs.(\ref{secorder})) form:
\begin{eqnarray}
&&\ddot{\sigma}+(d-1)\dot{\sigma}\dot{A}= V'(\sigma) ,    \label{eqmat}\\
&&\kappa \dot{\sigma}^2 = - 2 (d-2) \ddot{A} \left[ 1+ \lambda_0 \frac{(d-2)(d-4)}{4 m^2} \dot{A}^2 + \mu_0 \frac{(d-2)(d-6)}{m^4} \dot{A}^4 \right] ,    \label{eq2}
\end{eqnarray}
representing the matter and the scale factor equations, together with  the ``constraint" equation:  
\begin{eqnarray}
(d-1)(d-2)\dot{A}^2 \left( 1+ \lambda_0 \frac{(d-2)(d-4)}{4 m^2} \dot{A}^2 + \mu_0 \frac{(d-2)(d-6)}{m^4} \dot{A}^4 \right)
+ \kappa^2 V(\sigma) = \frac{\kappa^2\dot{\sigma}^2}{2}\label{constr}
\end{eqnarray}
which can be identified as describing the ``Hamiltonian" (i.e. of the $y$-momenta in our DWs case) constraint. Let us also mention that  the DWs equations (\ref{eqmat}), (\ref{eq2}) and (\ref{constr}) have been obtained from the  effective Lagrangian by replacing in the final results our specific \emph{gauge fixing} $f(y) = e^{-(d-1)A(y)}$, thus reproducing the initial form (\ref{dw}) of the DWs metrics, as it is explained in App.A.


\subsection{BPS-like first order DW's equations}

We are interested in the explicit construction of Lovelock DWs solutions relating two $AdS_d$ vacua $( \sigma_{k}^{*} , \Lambda_{eff}^{k} < 0 )$ for a family of (say, polynomial) potentials $V_{N}(\sigma)$ with few isolated extrema $\sigma_{k}^{*}$ $(k = 1,2,\ldots,N-1)$, i.e. $V'(\sigma_{k}^{*}) = 0$. As in the simplest case of the Einstein Gravity DWs (i.e. the $\lambda_0 = 0 = \mu_0$ limit of (\ref{gbl-action})) the only available exact analytic  DWs solutions can be obtained by the \emph{superpotential method} \cite{5}, \cite{6},  \cite{rg}, i.e. by  replacing the second order system (\ref{eqmat}), (\ref{eq2}) and (\ref{constr}) with an equivalent first order system of BPS-like equations. In order to generalize the well known results concerning the Einstein \cite{6},\cite{rg} and Gauss-Bonnet (see sect.7 of \cite{nmg} and ref.\cite{zee}) superpotential constructions to the cubic Lovelock case, we first introduce an auxiliary function $W(\sigma)$ of the matter field (called superpotential) such that $\dot{A} = - \kappa W(\sigma) / (d-2)$. We next realize that  the corresponding second order Lovelock DWs equations can be obtained from the following simple first order system:
\begin{equation}
\dot\sigma =\frac{2}{\kappa} W' \left( 1 + \lambda_0 \frac{(d-4)}{2(d-2)} \frac{\kappa^2 W^2}{m^2} + \mu_0 \frac{3(d-6)}{(d-2)^3} \frac{\kappa^4 W^4}{m^4} \right) ,\quad \dot{A} = - \frac{\kappa}{d-2} W(\sigma)          \label{sys}
\end{equation}
(where we have denoted $W'(\sigma) = \frac{dW}{d\sigma}$ and $\dot{\sigma}=\frac{d \sigma}{dy}$, etc.) together with the relation  between the matter potential and the superpotential:
\begin{eqnarray}
\kappa^2 V(\sigma) = 2 ( W' )^2 \left( 1 + \lambda_0 \frac{(d-4)}{2(d-2)} \frac{\kappa^2 W^2}{m^2} + \mu_0 \frac{3(d-6)}{(d-2)^3} \frac{\kappa^4 W^4}{m^4}\right)^2 -        \nonumber\\
- \left( \frac{d-1}{d-2} \right) \kappa^2 W^2 \left( 1 + \lambda_0 \frac{(d-4)}{4(d-2)} \frac{\kappa^2 W^2}{ m^2} + \mu_0 \frac{(d-6)}{(d-2)^3} \frac{\kappa^4 W^4}{m^4} \right)          \label{pot}
\end{eqnarray}
representing the constraint eq. (\ref{constr}).

For each given superpotential $W(\sigma)$ the solutions of the above first order system provide analytic DWs solutions of the cubic Lovelock Gravity coupled to scalar matter. For example in  three dimensions these are the DWs solutions of the cubic NMG Sinha's model \cite{sinha}. For $d=4$ the GB-terms are indeed not present in the DWs equations (due to the fact that now ${\mathcal{\chi}}_4 $ is a total derivative) and the cubic terms provide interesting new DWs solutions (and cosmological FRWs as well) containing Lovelock ($\mu$-dependent) corrections to the well known Einstein Gravity static flat DWs and Friedmann-Robertson-Walker (FRW) flat cosmological solutions. Finally, for $d\geq 5$,by construction, they represent the flat static  DWs of the Quasi-topological Gravity \cite{My_qtop},\cite{c_th} coupled to scalar matter of potential $V(\sigma)$.

It is worthwhile to mention that the reconstruction of the form of $W(\sigma)$ for a given matter potential $V(\sigma)$ by solving the non-linear first order eq. (\ref{pot}) is rather complicated problem and in general it provides more then one  solutions of different (global) properties\footnote{see refs. \cite {town1}, \cite{town2} for illuminating discussion concerning the case of the Einstein (super) Gravity DWs}. In the case of polynomial potentials $V(\s)$, one can find simple (finite order) polynomial solutions for the superpotential $W(\s)$ (by  series expansion method), which turns out to imply however certain relations between the the matter potential couplings and the gravity couplings. For example, assuming that  $V$ and $W$ are given by 
$V(\s) = \sum_{l = 0}^N g_l \, \s^l$ and  $W(\s) = \sum_{k = 0}^n \rho_k \, \s^k$,
one finds from (\ref{pot}) that $g_l = g_l ( g_i , \rho_j , \la, \m )$, together with the explicit form of the Superpotential parameters $\rho_k$ as functions of  all the couplings. Thus, besides the restrictions on its shape, reflecting the relations between the coefficients $\{g_l\}$, the matter couplings $\{g_l\}$ must be related to the   gravitational ones $\la$ and $\m$ in order to have consistent solutions of the constraint equation (\ref{pot}) and as a consequence well defined analytic Lovelock DWs. Four important remarks are now in order: 
 
 (1) Notice that once the relation between $\dot{A}(y)$ and the superpotential $W(\sigma(y))$ is assumed, the explicit form (\ref{sys}) of the $\dot\sigma$ in terms of W and its derivative $\frac{dW}{d\s}$ is a \emph{simple consequence} of one of the second order eqs.(\ref{eq2}) (when  $\dot\sigma\ne 0$); 
 
 (2) The second factor in $\dot\sigma$ equation, namely 
\begin{eqnarray}
 \C(W)= 1 + \lambda_0 \frac{(d-4)}{2(d-2)} \frac{\kappa^2 W^2}{m^2} + \mu_0 \frac{3(d-6)}{(d-2)^3} \frac{\kappa^4 W^4}{m^4}
\end{eqnarray} 
 could be easily recognized to coincide in the vacua  case (i.e. without matter, see for example \cite{My_qtop}, \cite{My_thol}) as the coefficient of the ``graviton's" kinetic terms, representing the contributions of the metrics fluctuations in the linearized (Gaussian) form of the Quasi-topological gravity action\footnote{The $\C(W)$ factor turns out to be simply related for \emph{odd} values of $d$ to the coefficient $c(\s)$  in front of the Weyl's tensor square appearing  in the trace anomaly in certain ``dual" $(d-1)$-dimensional $CFT$  and are known to play an important role in the description of RG flows \cite{rg}, as well as in the so called c/a-theorems \cite{x},\cite{2},\cite{c_th},\cite{My_thol}};  
 
 (3) Remembering that the explicit form of $\C(W)$ is very well known for all the dimensionally extended Lovelock gravity (higher then cubic)\cite{My_thol},\cite{bala},\cite{espanha},\cite{c_th} and also in the recent quartic generalizations of the Quasi-Topological gravity \cite{mann}, it becomes evident that the generalization of our cubic Lovelock first order system (\ref{sys}) as well as the Superpotential and the effective Lagrangian method  to these models is rather straightforward. This fact suggests that  the important problem concerning  the explicit construction of of \emph{flat Domain Walls} (and the related studies of c-theorems and RG flows) in a large family of generic extended Lovelock gravity coupled to matter matter, is now very well established and also not so much difficult to be realised. 
 
 (4) Let us also mention one more specific property of all the DWs solutions, based on the existence of first order BPS-like system (\ref{sys}), (and independently of the form of the superpotentials), namely  that the action can be always written as a total derivative:
\begin{eqnarray}
 S^{{\mathrm{eff}}}_{GBL} &=& -\frac{1}{\xi}\frac{2}{d-2} \int \! d^{d-1}x \, dy \; \frac{d}{dy}  \left[e^{(d-1)A(y)} \, (-\kappa W)^{d-1}\, a(W) \right],\\
 a(W)&=&\frac{\xi}{(-\kappa W)^{d-2}}\left( 1 + \frac{\lambda_0}{2m^2}\kappa^2 W^2 + \frac{3 \mu_0}{(d-2)^2 m^4}\kappa^4 W^4\right).\nonumber
 \end{eqnarray}
 Although the above result does not take into account the  contributions of the appropriate Gibbons-Hawking type of boundary terms (b.t.'s) (needed in order to make the variational problem  for the considered cubic extended Lovelock gravity coupled to matter well defined\footnote{The explicit form of these terms realised as derivatives of the extrinsic curvatures and Riemann tensors of near boundary $(d-1)$ -hypersurface for large family of (extended) Lovelock gravities are well known \cite{bterms1},\cite{bterms2}, but still they have to be appropriately "adapted" to the case of Quasi-topological gravity}), it is important to note that for odd $d$ again, they are simply related to the another (universal) trace anomaly coefficient $a(W)$ \cite{My_thol},\cite{2} playing an important role in the ``holographyc" calculations of entanglement entropies \cite{entrop},\cite{Myers-new}. Note that similarly to the $d=3$ New Massive Gravity case \cite{nmg},\cite{vicosa} the net effect of the  boundary terms  is in  the changes of the overall numerical coefficient in front of the action, but not in the W-dependence of the $a$-anomaly. This suggests that in the considered flat DWs case the Lovelock DWs tensions \cite{nmg}, \cite{6} can be realised in terms of (difference between) the a-anomaly values of the boundary and horizon CFTs and vice versa.

 Our final comment concerns the exact relation between the simplified cubic action (\ref{gbl-action}), we have used in the derivation of the second order Lovelock DWs equations, and the original Myers-Robinson's Quasi-topological gravity action (see eq.(4.25) of \cite{My_qtop}):  
 \begin{eqnarray}
S_{qtop} = &&\frac{1}{\kappa^2} \; \int \! d^d x \sqrt{-g} \Big\{ R - 2 \Lambda + \frac{ \lambda L^2}{(d-3)(d-4)}{\chi}_4\nonumber\\
 &&- \frac{8\mu (2d-3)L^4}{(d-6)(d-3)(3d^2-15d+16)}{\mathcal{Z}}_d +\kappa^2{ \mathcal{L}}_{\mathrm{m}} \Big\}   \label{qtop}
\end{eqnarray}
Let us first remind the explicit forms of the Gauss-Bonnet $ {\mathcal{\chi}}_4 $  and  ${\mathcal{Z}}_d $-invariants:  
\begin{eqnarray}
&&{\mathcal{\chi}}_4=R_{abce}R^{abce}-4R_{ab}R^{ab}+R^2,\nonumber\\
&&{\mathcal{Z}}_d = R_a{}^c{}_b{}^d \, R_c{}^e{}_d{}^f \, R_e{}^a{}_f{}^b + \frac{1}{(2d-3)(d-4)} \Big[ \frac{3(3d-8)}{8} R_{abcd} R^{abcd} R - 3(d-2) R_{abcd} R^{abc}{}_e R^{de} + \nonumber\\
&&+ 3d \, R_{abcd} R^{ac} R^{bd} + 6 (d-2) R_a{}^b R_b{}^c R_c{}^a - \frac{3 (3d - 4)}{2} R_a{}^b R_b{}^a R + \frac{3d}{8} R^3 \Big] \nonumber,
\end{eqnarray}
When written for an arbitrary conformally flat metrics: 
\begin{eqnarray}
&{\mathcal{Z}}_d = \frac{(d-3)(d(d+4)-4)(3d^2 - 15d +16)}{8(d-2)^3(d-1)^2(2d-3)} 
\left( R^3 - \frac{12d(d-1)}{d(d+4)-4} R R_{ab}R^{ab} + \frac{16(d-1)^2}{d(d+4)-4} R_{ab}R^{bc}R_c^a \right)\\
&{\mathcal{\chi}}_4=-\frac{4(d-3)}{(d-2)}\left(R^{ab}R_{ab}-\frac{d}{4(d-1)} R^2\right). \label{invgbl}
\end{eqnarray}
they take the exact from of the invariants used in our  action (\ref{gbl-action}). Then it becomes clear that for conformally flat metrics (and for $d\geq5$) our action (\ref{gbl-action}) coincides with the Myers-Robinson one if the corresponding Lovelock couplings are identified as follows:
\begin{eqnarray}
\frac{\lambda_0}{m^2}=-\frac{4\lambda L^2}{(d-2)(d-4)},\quad \quad \frac{\mu_0}{m^4}=-\frac{\mu L^4}{(d-6)(d-2)}\label{param}
\end{eqnarray}
and the cosmological constant $\Lambda$ is given by the vacuum value of our matter potential: $\kappa^2 V(\sigma^*)=2\Lambda = -(d-1)(d-2)/L^2$. 
Notice that in six dimensions the cubic terms do not contribute to the DWs solutions, which now  have the same form as in the pure GB gravity case, i.e. $\mu=0$ case, due to the well known fact that the cubic Lovelock invariants in $d=6$ have the form of  a total derivative.


\section{On the vacua space of  extended Lovelock Gravity with matter}

\setcounter{equation}{0}

The AdS$_d$ -type vacua solutions  $(\sigma_{k}^{*} \, , \, \Lambda_{\mathrm{eff}}^{k} < 0)$ of eqs. (\ref{sys}) are defined  by $\dot{\s}  = 0$, being extrema of the matter potential $V'(\sigma_{k}^{*})=0$ and reproducing  the geometry of an AdS$_d$ space of radius $L_k^2 = - (d-1)(d-2) / 2 \Lambda_{\mathrm{eff}}^k$ in Poincar\'e coordinates (see for example ref. \cite{nmg}). The ``effective" AdS$_d$ scale $L^2_k$ is related to the superpotential's vacuum values  by  $L_k^{-2} = \ka W^2(\s_k^*) / (d-2)^2 $,
as one can see  from the expression for the curvature:
\begin{equation}
R = - 2 (d-1)\left[ \frac{d}{2} \dot{A}^2  + \ddot{A} \right] = - 2(d-1) \left[ \frac{d}{2(d-2)^2} \,\kappa^2 W^2(\s) - \frac{\kappa}{d-2} W'(\s) \, \dot{\s} \right] \; ,  \label{curvature}
\end{equation}  
 i.e. we get indeed  $R(\s_k^*) = - d(d-1)/L_k^2 = - d ( d-1) W^2(\s_k^*) / (d-2)^2$. Evaluated at each of the vacua, the first term in the r.h.s. of eq. (\ref{pot}) vanishes and it takes the form of the following cubic ``vacua" equation: 
\begin{equation}
h_k =  f_k ( 1 - \la f_k - \m f_k^2 ) \;, \label{cubic}
\end{equation}
where we have introduced as in \cite{My_qtop} the ``bare" $h_k \equiv  L^2/L_{0k}^2 = - L^2 \, V(\s^*_k) / (d-1)(d-2) $ and the effective $f_k \equiv L^2/L_k^2 = L^2 \dot{A}^2(\s^*_k) = \ka^2 L^2 W^2(\s^*_k) / (d-2)^2 \;$ ($f_k>0$) scales, correspondingly\footnote{Each extremum $h_k$ of the matter potential $V(\s)$ can be seen as a ``bare'' cosmological  constant $\Lambda_k^{0}=-(d-1)(d-2)/ L_{0k}^2$, appearing explicitly in the Lagrangian.}. The equation above shows that, for each one of these bare cosmological constants, we can find up to three different \textit{effective} cosmological constants $\Lambda_{\mathrm{eff}}^k$ related to the ``effective" radii $L_k$.

\subsection{Stability conditions}

We are interested in those of the vacua of the gravity-matter model (\ref{qtop}) that  satisfy the following \emph{three} conditions:  

The \emph{first} is $h_k > 0$, or $V(\sigma^*_k) < 0$, meaning that the related bare cosmological constant is always negative. It is usually imposed in order to have consistent (perturbatively stable) limits to the Einstein-Hilbert (EH) gravity of negative cosmological constant.

The \emph{second} is  the \emph{stability} (causality/positive energy) requirement, selecting only those of the  solutions $f_k>0$ of eq. (\ref{cubic}) with
\begin{eqnarray}
  \C(f_k) \equiv \partial_{f_k} h_k (f_k) = 1 -2 \la f - 3 \m f^2>0  \label{stab}
\end{eqnarray}  
 This condition thus  excludes all the $f_k$'s that lead to the wrong sign (i.e. ``ghosts") of the graviton's kinetic terms in the corresponding ``linearized" (Gaussian) form of the Quasi-topological Gravity action (\ref{qtop})\footnote{valid for $d>4$ only.}
(as demonstrated in refs.\cite{My_qtop}, \cite{My_thol}). 

The \emph{third} is the well known Breitenlohner-Freedman (BF) unitarity condition \cite{BF} for massive scalar field in AdS$_d$ background:
\begin{eqnarray}
-\frac{(d-1)^2}{4 L_k^2}\le m_{\sigma}^2(\sigma_{k}^*)=V''(\sigma_{k}^*)\label{BFred}
\end{eqnarray}
It also ensures the stability of the gravity-matter model vacua $(\sigma_{k}^{*} \, , \, \Lambda_{\mathrm{eff}}^{k} < 0)$  with respect now to the linear fluctuations of the scalar field, thus providing a consistent (positive norms, i.e unitarity)  quantization of this scalar field on AdS$_{d}$ background of given $\Lambda_{{\mathrm{eff}}}$.
 
 We next observe that according to  eq. (\ref{sys}) and (\ref{pot}) the extrema of $V(\s)$ are given by all the (real) solutions of the following equation:
\begin{eqnarray}
V'(\sigma_k^*)=0=\frac{2W'}{\kappa^2}\C(f_k)\mathcal{F}(\sigma)\Big|_{\sigma_k^*}\label{extrem}
\end{eqnarray}
where the first two factors are the same as that in the $\dot{\s}$ eq.(\ref{sys}), and the new one 
\begin{eqnarray}
\mathcal{F}(\sigma)=2W''\C(f_k)-\frac{8\lambda L^2}{(d-2)^2}W'^2\kappa^2W\left(1+\frac{3\mu}{\lambda(d-2)^2}L^2\kappa^2W^2\right)-\left(\frac{d-1}{d-2}\right)\kappa^2W \nonumber
\end{eqnarray}
is representing those of the $V$ extrema that are not described by the $I^{st}$ order system. Therefore we have to distinguish the following  three types of vacua ($\sigma_k^*$, $f_k$), originated from the three distinct factors $W'(\s)$, $\C(\s)$ and $\mathcal{F}(\sigma)$ present in $V'(\s)$: 
 
(\textit{a})  The \emph{first} one is given by the extrema of the superpotential $W'(\s_k^*) = 0$ with $W(\s_k^*)\neq0$\footnote{Notice that the vacuum with $W(\s_k^*)=0$ represents a Minkowski space-time of $\Lambda^{eff}=0$  and they are not further considered  since we are interested here in the vacua of $AdS$-type only.} giving rise to both positive (``physical") or negative (``ghost"-like) values of $C_0(f_k)$;

(\textit{b}) the vacua of \emph{second} type (called \textit{topological}) are defined by the real solutions of the equation $\C(f_{top}) = 0$;

(\textit{c}) the \emph{third} type of vacua are given by the solutions of the (non-linear) equation $\mathcal{F}(\sigma_k^*)=0$ and they are \emph{not} of BPS-type differently from the first two types of vacua defined above\footnote{They are not representing in fact the vacua of the corresponding supergravity coupled to chiral matter models, which are in the origin of the $Ist$ order equations as the conditions of the existence of constant Killing spinors (see for example \cite{6},\cite{town1},\cite{town2},\cite{rg})}. There exist indications that such ``non-supersymmetric" vacua are \emph{unstable} (see refs.\cite{town2},\cite{5},\cite{town1}). They will appear however in our discussion of the implications of the ``extended" BF-conditions (\ref{news}) on the shape of the matter potential in Sect.4.4 (see Fig.2b) below.

Hence an important consequence of the superpotential description (\ref{sys}) of the Quasi-topological Gravity coupled to matter (valid for vacua and flat DWs only) is that one can always find an appropriated range of values of the gravitational couplings $\la$ and $\mu$ (not both negative), such that  at least one of the vacua of the considered  gravity-matter model (\ref{qtop}) is of topological nature, i.e. with $C_0 = $0. The problem addressed in this section concerns the derivation of the complete set of conditions that single out the physical vacua ($f_k$, $C_0(f_k)>0$) among all the  AdS$_d$-type extrema of $V(\sigma)$, taking into  account the fact that one or two of the vacua of the model (\ref{qtop}) must be topological ones. Let us first demonstrate that the proper existence of topological vacua:

 $i$) introduces natural smallest or largest AdS$_d$ scale(s) that can be chosen as a fundamental scale $L^2=L^2_{0,top}$ present in the action (\ref{qtop});
 
 $ii$) it determines two particular families of models  corresponding to specific relations between the  Lovelock couplings $\lambda$ e $\mu$.

\vspace{0.5cm}

\textit{The Gauss-Bonnet-matter example.} The  description of the restrictions on the gravitational couplings, as well as those on the physical vacua scales, in the simpler GB case (i.e. $\mu=0$) consists in selecting the set of solutions $f_k$ of GB vacua equation (\ref{cubic}): $h = f ( 1 - \la f)$, that satisfy the stability condition $\C = 1 - 2 \la f>0$. If $\la < 0$, no topological vacuum exists at all and as a consequence all the vacua are physical, i.e. $h(f), \C(f) > 0$. In the case $\la > 0$ we find that $h(f) > 0$ held  for $0 < f < 1/\la$, while   $\C(f) > 0$ takes place for $f < 1/2\la$ only. Hence the physical region is  defined by $0 < f < 1/ 2 \la$ and the (only) topological vacuum is given by  $f_{top} = 1/2\la$ and $h(f_{top}) = 1/4 \la$. Notice that $f_{top}$ is the boundary of the region of the physical vacua, i.e. one always has $f_{phys} < f_{top}$, and therefore $f_{top} \sim L_{top}^{-2}$ represents the smallest AdS$_d$ radius (scale). It is customary to make the normalization $h(f_{max}) = 1$, where $f_{max}$ is the greatest scale physical vacuum. This leads to the restriction $1/ 4 \la = h(f_{top}) > 1$, or $\la < 1/4$. However, one can make an alternative normalization $h(f_{top}) = 1$, which \textit{fixes} $\la = 1/4$. With this normalization, there is an upper boundary to the physical bare cosmological constants: $h(f_{phys}) < 1$. Although apparently very distinct, both normalizations are in fact equivalent, since the term appearing in the Lagrangian, $L^2 \la$, is the same in both cases.

\subsection{Classification of Lovelock-matter vacua}

We next analyse the  restrictions that the  physical vacua  of Quasi-topological Gravity coupled to matter  (\ref{qtop}) should satisfy. Now  the ``topological" vacua equation $C_0(f_{top})=0$  is of second order and therefore its two solutions are given by:
\begin{equation}
f_\pm = - \frac{1}{3\m}( \la \mp \sqrt{  \la^2 + 3 \m  } )  \; . \label{f top mu and la}
\end{equation}
The sign of the second derivative $h''(f_\mp)$ shows that $f_+$ is a local maximum of $h(f)$, while $f_-$ is a local minimum\footnote{Notice that even if we sometimes consider $h(f)$ or $\C(f)$, as continuous functions, it must be kept in mind that the set of vacua $\{ f_k \}$ is in fact discrete.}.
By substituting (\ref{f top mu and la}) in eq. (\ref{cubic}) we find the explicit form of the $h_{top}^{\pm}$ for the corresponding topological vacua
\begin{equation}
h_\pm = h(f_\pm) = \frac{1}{27\m^2}\left[ - \la (2 \la^2 + 9 \m) \pm 2 (\la^2 + 3 \m)^{3/2} \right]\; . \label{h top la and mu}
\end{equation}
These equations can be further solved for the Lovelock coupling $\mu=\mu(h_{\pm},\lambda)$ as follows\footnote{As we have mentioned, we are excluding here the case when both $\la$ and $\mu$ are negative.}:
\begin{eqnarray*}
&&\m_+ (\la) = \frac{1}{27h_+^2}\left[ 2 - 9 \la \, h_+ + 2 ( 1 - 3 \la \, h_+ )^{3/2} \right], \ \textrm{if } \la<0,\\
&&\m_\pm (\la) = \frac{1}{27h_{\pm}^2}\left[ 2 - 9 \la \, h_\pm \pm 2 ( 1 - 3 \la \, h_\pm )^{3/2} \right],\ \textrm{if } 0<\la<1/3,
\end{eqnarray*}
thus defining for each fixed value of $0<\lambda<1/3$ two different values $\m_{\pm}(\lambda)$ whenever topological vacua do exist. 
One can easily verify that $h_-(\m_-) > 0$ and $h_+(\m_\pm) > 0$ for all $\la$, which suggests that one can  always choose the fundamental  scale as $L^2=L_{0+}^2(\mu_+)=L_{0-}^2(\m_-)$  by  normalizing the ``bare" topological vacuum as $h_+(\m_+, \la ) = h_-(\m_-, \la) = 1 $, i.e. taking the smallest topological scale of the $\mu_+$ model, and the largest one of the $\mu_-$ model. Therefore, for each $\la$, there are two distinct gravitational models, corresponding to the two different forms of the Lovelock coupling $\mu$ as a function of $\lambda$:
 \begin{equation}
\m_\pm (\la) = \frac{1}{ 27}\left( 2 - 9 \la \pm 2 ( 1 - 3 \la  )^{3/2} \right)   \; , \label{mu pm}
\end{equation}
and having equal fundamental scales. As expected, these are exactly the two curves $\m_-(\lambda) < \m_+(\lambda)$ where two of the roots of the cubic ``vacua" eq. (\ref{cubic}) do coincide (see fig.1 of the Myers-Robinson paper \cite{My_qtop}). It can be also shown that $\m_- < 0$ for all $\la < 1/3$, $\m_+ > 0$ for $\la < 1/4$, and $\m_+ < 0$ for $1/4 < \la < 1/3$. At $\la = 1/3$ and $\mu= -1/27$ both curves coincide and terminate,  describing the degenerate case when all the three  roots of vacua eq. (\ref{cubic}) are equal. Let us also mention that the \emph{two} particular cases $\la = 1/4$, $\m_+ = 0$ (which corresponds to GB gravity) and the usual EH gravity case with $\la = 0$, $\m_- = 0$ in fact belongs to a different family of $\mu$ models: the $\m_+$ and the $\mu_-$ respectively.

In order to make transparent the ``physical" restrictions on the $f_k$'s  for each fixed  value of $\la < 1/3$ (in both of the $\mu_{\pm}$ models), we represent all the vacua of the gravity-matter model (\ref{qtop}) as a set of points $(h_k, f_k)$ on the $h$-$f$ plane,  belonging to one of the two cubic curves $h(\la, \m_\pm ; f)$.

\begin{figure}[ht]
\centering
\subfigure[]{
\includegraphics[scale=0.55]{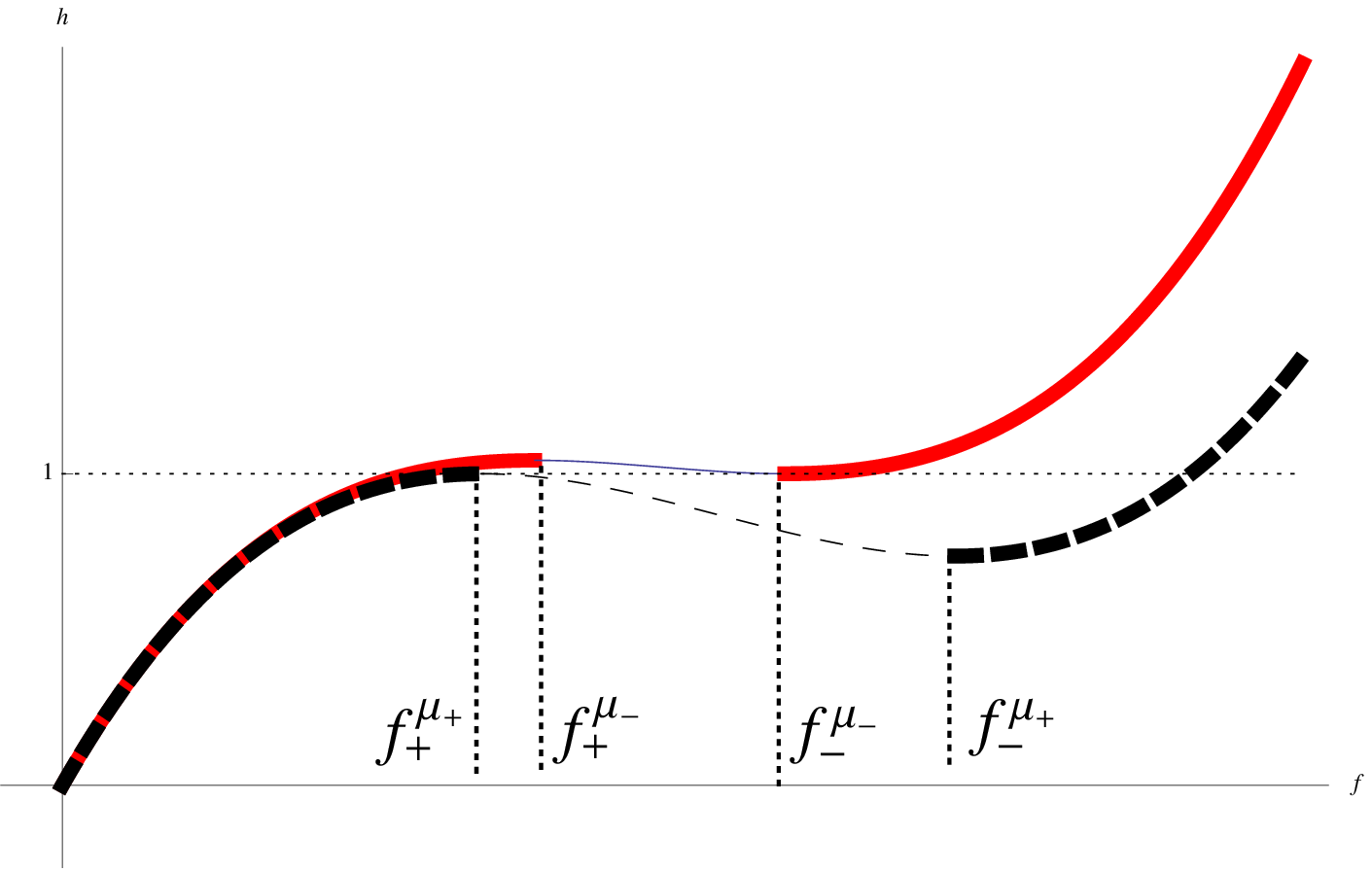}
\label{fig:1a}
}
\subfigure[]{
\includegraphics[scale=0.55]{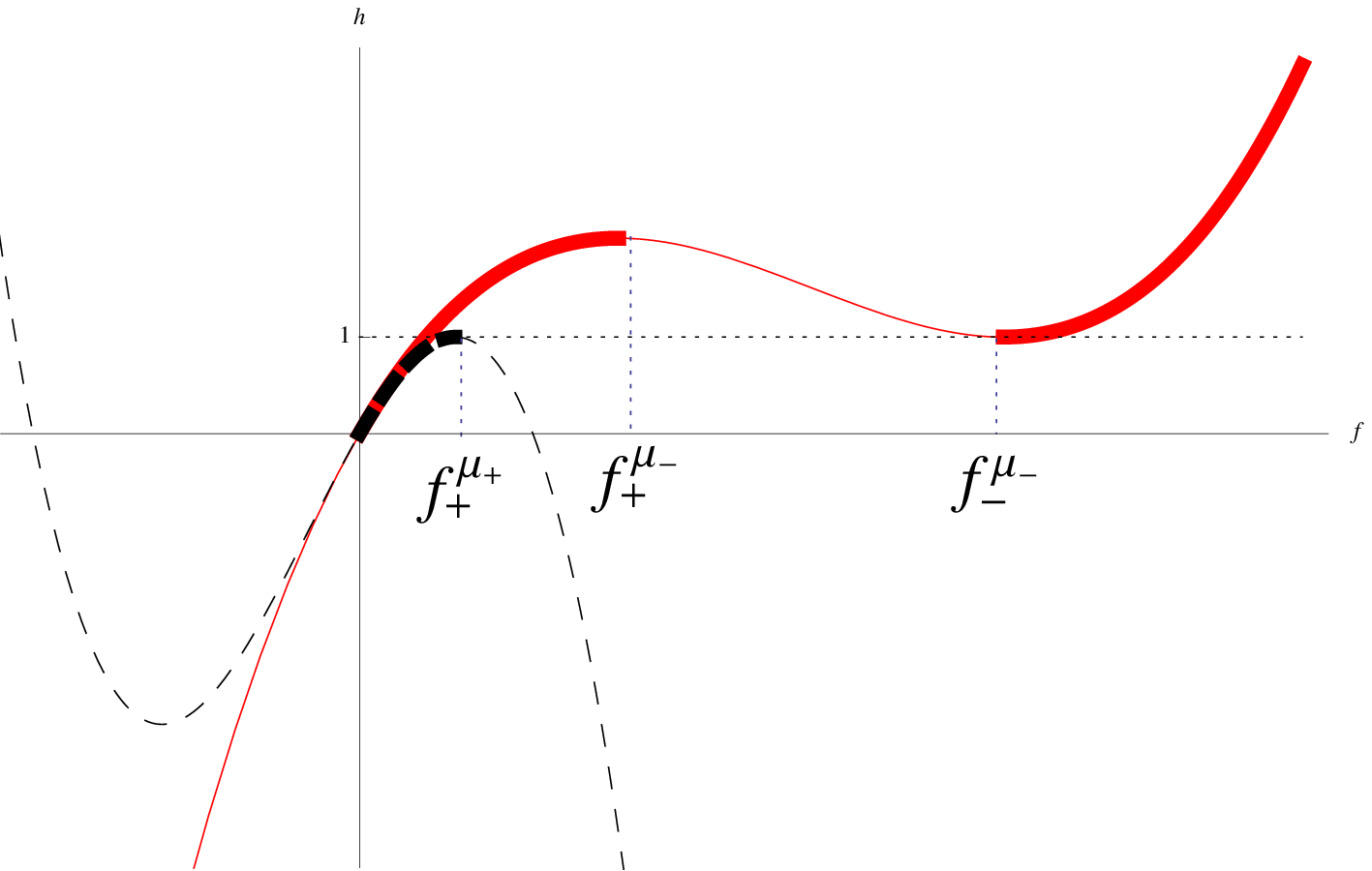}
\label{fig:1b}
}
\label{fig:1}
\caption{The curves $h(f)$ for the models $\mu_{\pm}$, after the normalization: $(a)$ the case $\frac{8}{27}<\lambda<\frac{1}{3}$, $(b)$ The region represented here is $0<\lambda<\frac{1}{4}$. The red continues curves correspond to $\mu_-$ models, while the black dashed ones - to the $\mu_+$ models. The physical regions present in both curves are the thick ones.}
\end{figure}


\textit{The $\mu_+$ model:}
It is clear that the physical regions for the model corresponding to the  curve $h(\la, \m_+ ; f) \equiv h(f)$ are limited by its  extrema (corresponding to the topological vacua): 
\begin{eqnarray}
&& h_+ = h(f_+) = 1\, , \;\; h_- = h(f_-) =  \frac{4 - 45 \la + 108 \la^2 - 4 (1 - 3\la)^{3/2}}{27 \la (1 - 4 \la )^2}, \\
&& {\mathrm{with}}  \;\;  f_+ =\frac{1}{\la} ( 1 - \sqrt{1 - 3\la} ) \, , \;\; f_- = \frac{1}{3\la(1 - 4\la)}\left( 1 - 6 \la - \sqrt{1 - 3  \la} \right).
\end{eqnarray}

Notice that for $\la < 1/4$, the minimum is negative: $f_- < 0$ and therefore the physical conditions $h, \C > 0$ are  fulfilled only when  $0 < f_{phys} < f_+$ (see Fig.\ref{fig:1b}). This case is qualitatively similar to the physical region of GB gravity, i.e. we have only  one topological vacua $f = f_+$ that defines the  \textit{minimal} scale $L_+^2 \sim f_+^{-1}$ of the model. 
 
We next consider the interval $1/4 < \la < 8/27$ where both  $f_{\pm} > 0$, but now it turns out that $h_-$ is negative, thus violating the AdS condition: $h > 0$. It is nevertheless possible to construct stable, analytic DWs if the effective cosmological constant is greater than a certain value, i.e. $f_{phys} > f_0$, where $f_0 = - \la / 2 \m_+ + \sqrt{ \left( \la^2 + 4 \m_+ \right)/4 \m_+^2}$ is the greatest root of $h(f) = 0$ equation. The typical GB-like physical region for $0 < f_{phys} < f_+$ is also present in this case.
 
Finally, in the last region   $8/27 < \la < 1/3$ we find that both $f_\pm$ are positive, as well as both $h_\pm$. We realize that in the region between the two extrema  $f_+ < f_{phys} < f_-$ we always have $\C < 0$, i.e. all the vacua solutions belonging to this interval  have to be excluded since they are ``ghost"-like. As a consequence the physical region is divided into two subregions of qualitatively different types: one is the usual GB-like region given by $0 < f_{phys} < f_+$, while the  other one, $f_- < f_{phys}$, is of a new type, having  no upper limit on the values of the effective cosmological constants (see Fig.\ref{fig:1a}). In this case the topological vacua $f_-$ introduces a natural \emph{ maximal scale}  $L_-^2 \sim f_-^{-1}$, thus limiting the possible values of the physical vacua scales from above.

\textit{The $\mu_-$ model:} It is symbolically represented by the curve $h(\la, \m_-; f) \equiv \tilde{h}(f)$. It has  a rather simpler vacua structure, similar to the one of the $\mu_+$ model for  $8/27 < \la < 1/3$ described above, due to the fact that now $\tilde{h}_- > 0$ for all $\la > 0$. Its extrema have a  slightly  different form:
\begin{eqnarray}
&& \tilde{h}_+ = \tilde{h}(\tilde{f}_+) = \frac{4 - 45 \la + 108 \la^2 + 4 (1 - 3\la)^{3/2}} { 27 \la (1 - 4 \la )^2} \, , \;\; \tilde{h}_- = \tilde{h}(\tilde{f}_-) = 1 ,  \label{h tilde} \\
&& {\mathrm{with}}  \;\;  \tilde{f}_+ = \frac{1 - 6 \la + \sqrt{1 - 3  \la} }{  3\la(1 - 4 \la)} \, , \;\; \tilde{f}_- = \frac{ 1 + \sqrt{1 - 3\la} }{  \la} ,
\end{eqnarray}
and they indeed represent the limits of the corresponding physical regions (see Fig.\ref{fig:1a}). Thus, its  GB-like (i.e. \emph{minimal} scale) physical region is given by $0 < f_{phys} < \tilde{f}_+$, while in the region $\tilde{f}_- < f_{phys}$ the physical vacua scales are restricted by the corresponding $\mu_-$ model \emph{maximal} scale $L_{-}^2(\m_-)$.

A short comment about our specific choice  $L^2 = L_{0+}^2(\m_+) = L_{0-}^2(\m_-)$ of the \emph{fundamental} scale for the Quasi-Topological Gravity coupled to matter (\ref{qtop}) is now in order. It is important to emphasize that although such choice is rather natural and general (in the cases when topological vacua do exist), we can equivalently use all the other bare or effective topological scales. For example, the \emph{minimal bare} topological scale for the $\m_-$ model,  $\tilde{L}_{0+}^2$, can be easily rewritten as $\tilde{L}_{0+}^2 = \tilde{L}_{0-}^2 / \tilde{h}_+$ with $\tilde{h}_+$ given by (\ref{h tilde}). Similarly, each one of the  effective scales, for example $f_+ = L^2/L_+^2 = L_{0+}^2 / L_+^2$, can be expressed, say, for the $\m_+$ model as  $L_+^2 = (1 + \sqrt{1 - 3 \la}) L_{0+}^2/ 3$, i.e. in the terms of the corresponding \emph{bare} scale. Completely analogous relations between $L_{0+}^2(\m_+)$, $L_-^2(\m_-)$ and all the other topological scales, both bare and effective, for $\m_+$ and $\m_-$ are easily found, making the chosen normalization equivalent to any other.  Let us also mention that all the restrictions on the physical vacua we have derived in this section  take a rather simple form when  rewritten in terms of the corresponding effective scales. For example, in the GB-like regions, one has $L^2_{phys} > L_+^2$, while  the physical restrictions in the \emph{maximal} scale regions reads now as $L^2_{phys} < L_-^2$ .


\subsection{Consequences of BF and unitarity conditions}

In order to derive the additional restrictions on the physical vacua (and on the shapes of  $W(\sigma)$ and $V(\sigma)$) imposed by the BF-condition (\ref{BFred}), it is convenient to rewrite the ``effective" vacua scalar field masses $m_{\sigma}^{2}(\sigma_{k}^{*})$ in the following suggestive form:
\begin{eqnarray}
m_{\sigma}^{2}(\sigma_{k}^{*})=\frac{\kappa^2W_{A}^2}{(d-2)^2}s_{k}(s_{k}-d+1)=\frac{1}{L_k^2}s_{k}(s_{k}-d+1)\label{mas}
\end{eqnarray}
which is valid in the both cases of  ``physical" (or ``ghost"-like) $\sigma_k^*$ and for the topological $\sigma_{top}^{\pm}$ vacua, all denoted by  $\sigma_{A}^{*}$ in eq.(\ref{mas}) above. It is obtained from eq. (\ref{pot}) by evaluating  the $V''(\sigma)$ values at the corresponding vacua (i.e. on the $V$'s extrema).  Assuming that $W_{k}\neq0$, we have introduced as new parameters the so called ``critical exponents":
\begin{eqnarray}
&s_a = 2 (d-2) \frac{W''_a}{\kappa^2W_a} \left[ 1 - \frac{2 \la L^2}{(d-2)^2} \kappa^2W_a^2 - \frac{3 \m L^4}{(d-2)^4}\kappa^4 W_a^4 \right]
    = 2 (d-2) \frac{W''_a}{\kappa^2W_a} \left(1 - \frac{L_+^2}{ L_a^2} \right)\left(1 -\frac{ L_-^2 }{ L_a^2} \right),\nonumber\\
&s_{top}^\pm = - \frac{8}{(d-2)} W_{\pm}'^2 \left[  \la L^2 + \frac{3  \m L^4}{(d-2)^2}\kappa^2 W_{\pm}^2 \right] = - \frac{4 L_\pm^2}{(d-2)} W_{\pm}'^2 \left[ 1 - \frac{L_\mp^2}{L_\pm^2} \right]\label{criti}
\end{eqnarray}
related (according to $AdS/CFT$ correspondence \cite{witt}) to the scaling dimensions:
\begin{eqnarray}
\Delta_{k}^{\pm}=d-1-s_{k}^{\pm}=\frac{d-1}{2}\pm\sqrt{\frac{(d-1)^2}{4}+m^2(\sigma_{k}^*)L_{k}^2}\label{sdime}
\end{eqnarray}
of the ``dual" fields $\Phi_{\sigma}(x_i)$ in the corresponding $CFT_{d-1}$. It is worthwhile to mention that the parameters $s_{k}^{\pm}$  in fact determine the asymptotic behavior of the matter field $\sigma(y)$ (see App. B for more details):
\begin{eqnarray}
\sigma(y)\stackrel{y\rightarrow\infty}{\approx}\sigma_{k}^* +const.\, e^{-
s_k\frac{y}{L_k}}, \quad \label{asymp}
\end{eqnarray}
Thus the values of ``critical exponents" $s_k\neq 0$ provide an important additional information about the boundary conditions (b.c.'s) for the corresponding DW's solutions of the model as one can easily verify by considering the near-boundary/horizon's approximation  of eqs.(\ref{sys}). The same arguments applied to the degenerate case $s_k=0$ of say two coinciding vacua, i.e when $W'$ has double zero at some $\sigma_k^*$:
 \begin{eqnarray}
\dot{\sigma}\approx \frac{1}{\rho_k} (\sigma - \sigma_k^*)^{2}\kappa W_{k},\quad\quad
\sigma(y)\stackrel{y\rightarrow\infty}{\approx}\sigma_{k}^* + \Big(\frac{y}{\rho_k L_k} \Big )^{-1}\label{asympde}
\end{eqnarray}
lead as expected to qualitatively different power-like b.c.'s  for $\sigma$ instead of the exponential decay (\ref{asymp}) specific for the case of non-degenerate simple zeros of $W'$. Since for $m^2(\sigma_{k}^*)\ne 0$, the two roots  $\Delta_k^{\pm}$ give rise to two different $s_k^{\pm}$, they are therefore corresponding to different b.c.'s for $\s$  and as argued in ref. \cite {witku} they lead to different states and to different quantizations of the ``boundary" $CFT_{d-1}$.

 We next realise that  the  BF-condition (\ref{BFred}) are in fact  automatically satisfied for all the (real) values of $s_{k}$. However for the $AdS/CFT$'s applications an important further restriction is the \emph{unitarity} of the dual $CFT_{d-1}$ (conjectured to exists for each $AdS_{d}$ vacua), which requires the \emph{positivity} of the scaling dimensions $\Delta_{k}^{\pm}>0$ or equivalently $s_{k}<d-1$. Notice that the values of $s_k^{-}>d-1$ leads to $\Delta_k^{-}<0$ and therefore to \emph{non-unitary} CFT's. Although  both  scaling dimensions (under some restrictions) have consistent but distinct $AdS/CFT$ applications, we are  further considering only the  $\Delta_k^+$  root of quadratic eq. (\ref{mas}) that is known to be relevant for the description of the off-critical behavior of the corresponding CFT perturbed by $\Phi_{\sigma}(x_i)$ . We shall also impose the condition that a part of the operators $\Phi_{\sigma}(x_i)$ (those ones corresponding to UV critical points) to be \emph{relevant}, i.e. $0<\Delta_{UV}^{+}<d-1$. For the $\Delta_k^+$ root we have chosen this condition reads as $\frac{d-1}{2}<\Delta_{UV}^+<d-1$ or equivalently the critical exponents should satisfy: $0<s_{UV}^+<\frac{d-1}{2}$, which takes place only for \emph{negative} $m^2_{UV}$ within the frameworks of BF condition. Note that the  positive values of  $m^2_{UV}>0$ corresponds to $\Delta_{UV}^+>d-1$, (i.e negative $s_{UV}^+<0$), which are indeed  discarded as irrelevant operators. As we shall demonstrate in Sect.4.4. below, these two $CFT_{d-1}$ motivated additional requirements:
\begin{eqnarray}
-\frac{(d-1)^2}{4}\le m_{\sigma}^2(\sigma_{UV}^*)L_{UV}^2<0, \quad \quad\quad  0<s_{UV}\le \frac{d-1}{2}  \label{news}
\end{eqnarray}
(i.e. the stronger form of the BF-condition) together with the other physical restrictions on values of $AdS$ scales $L_k$ derived in this section, lead to important qualitative changes in the shape of the matter potential, compared to the case when $s_{UV}>d-1$. 

\section{Domain Walls }

\setcounter{equation}{0}

The $AdS_d$ type vacua solutions $(\sigma_k,L_k,s_k)$ of the considered gravity-matter model (\ref{qtop}) that satisfy  all the physical conditions derived in Sect.3, provide a set of admissible b.c.'s (at $y \rightarrow\pm\infty$) for the stable Lovelock DW's we are interested in. By definition such DW's relate two neighbouring vacua, representing the (null) horizons or/and the  boundaries of certain asymptotically $AdS_d$ space-time. Since all the restrictions are in terms of the superpotential vacuum values $W_k$ and $W''_k$ our problem consists in finding  an appropriate  $W(\s)$ that generates $V(\s)$ with at least two consecutive physical vacua and eventually (depending on the values of $\la$ and $\m$) few topological ones. We next consider the quartic superpotential $W(\s) = - B [ ( \s^2 - x_0 )^2 + D ]$ of inverted ``double-well" type which for $B$, $D$ and $x_0$ all positives , allows an explicit analytic constructions of physical DW's for all the permitted values of the gravitational couplings. Few particular examples corresponding to the case of $B<0$, i.e. to the standard Higgs-like superpotential ,are also admitting  physical DWs but now in the regions of maximal scale (see Fig.1). Due to the reflection $Z_2$-symmetry  $W(-\s) = W(\s)$ the above quartic $W(\s)$ has an advantage to permit an easier integration (and a rather simple form) for the scale factor $\exp(2A(\s))$ compared to the case of cubic superpotentials\footnote{The explicit constructions of Lovelock DWs for a family of \emph{quadratic} superpotentials are presented in App.B bellow. However  they are not providing an example for \emph{stable physical} DWs, since  as it turns out they are always relating either a physical vacua to the topological one or two topological vacua} . We can therefore restrict our analysis  to the case  $\s > 0$ only. The sign of $BD$ is further determined by an extra condition  $W(\s) \leq 0$ we have imposed in order to ensure that $\dot{A}(y)$ does not change it sign. This choice is fixing  the horizon at $y \rightarrow - \infty$ and the the boundary to be at $y \rightarrow + \infty$, thus excluding all the cases of $(a)AdS_d$ spaces  having two horizons or two boundaries.

\subsection{Lovelock vacua for quartic Superpotential}
The extrema of the superpotential for $\s\geq 0$ (i.e.  $W'(\s) = 0$) denoted by  $\s_{{\mathrm{IR}}} = 0$ and $\s_{{\mathrm{UV}}} = \sqrt{x_0}$ are  candidates for representing the physical vacua. We next realize the $W(\s)$ parameters in terms of the vacua scales\footnote{in this section and in the App. B we are fixing for simplicity $\kappa=1$.} introduced in Sect.3:
$$W_{{\mathrm{IR}}} = - B( x_0^2 + D ) \; , \;\; f_{{\mathrm{IR}}} = \frac{L^2 B^2( x_0^2 + D )^2}{(d-2)^2}  \; ; \;\; W_{{\mathrm{UV}}} = - B D \; , \;\; f_{{\mathrm{UV}}} = \frac{L^2 B^2 D^2 }{ (d-2)^2} \; , $$
where $W_{{\mathrm{IR}}} = W(\s_{{\mathrm{IR}}})$, etc. and $f_\ir = L^2/L_\ir^2 $. Here $L^2$ is given by the normalization established in Sect.3., i.e. $L^2 = L^2_{0+} = \tilde{L}^2_{0-}$.
It follows that
\begin{equation}
B x_0^2= \frac{(d-2)}{L} \left( \sqrt{f_{{\mathrm{IR}}}} - \sqrt{f_{{\mathrm{UV}}}} \right)  \; , \;\;\;\; B D = \frac{(d-2)}{L} \sqrt{f_{{\mathrm{UV}}}}  \; , \label{B and BD}
\end{equation}
and therefore  the sign of $B$ determines which of the effective vacua radii (i.e. the scales $L_k$) $f_\ir$ and $f_\uv$ is the greatest one. Notice that the conditions on the $f_{phys}$ derived in Section 3, are now easily transformed in certain restrictions on the parameters of $W(\s)$. For example, taking $B > 0$ and considering the $\m_+$ model in the GB-like region we find that 
$$L_{0+}^2 B^2 D^2 <   \frac{(d-2)^2 }{\la} (1 - \sqrt{ 1 -3 \la})$$
must be satisfied in order to have two physical vacua.
Changing the sign of $B < 0$, i.e. for the case the \emph{standard "double-well"} superpotential  and  considering now the $\m_-$ model in the region with a \emph{ maximal scale}  we conclude that 
$$\tilde{L}_{0-}^2 B^2(x_0^2  + D)^2 >  \frac{(d-2)^2}{\la}  (1 + \sqrt{ 1 - 3\la}) $$
is ensuring that the corresponding vacua $\s_{{\mathrm{IR}}}$ and $\s_{{\mathrm{UV}}}$ are both physical.

We next describe the properties and the conditions on the topological vacua for our quartic superpotential, remembering that we can have only one such vacua for $\m>0$, two for $\m<0$ and $0<\la < 1/3$ and no one when both $\la$ and $\m$ are negative, as one can verify from its proper definition $\C(W_{top})=0$. Then the topological vacua $\s_{top}^\pm$ are given by all the real solutions of the following equation:
\begin{equation}
\left(\frac{ (\s_{top}^\pm)^2 }{x_0} - 1 \right)^2 = \frac{ \sqrt{f_\pm} - \sqrt{f_{{\mathrm{UV}}}} }{   \sqrt{f_{{\mathrm{IR}}}} - \sqrt{f_{{\mathrm{UV}}}}},   \label{sigma top eq}
\end{equation}
where we have used eqs.(\ref{B and BD}) and  also the notation $(W_{top}^\pm)^2 = (d-2)^2 f_\pm / L^2 > 0$ was adopted. The sign of the denominator is the sign of $B$, hence the conditions for the r.h.s. of the last equation to be positive are as follows : in the case  $B > 0$ we have to impose  $f_{{\mathrm{UV}}} < f_{{\mathrm{IR}}}$ and $f_{{\mathrm{UV}}} < f_\pm$, while in the case  $B < 0$ we get  $f_{{\mathrm{IR}}} < f_{{\mathrm{UV}}}$ and $f_{{\mathrm{UV}}} > f_\pm$. Then  it is easy to show that if $B > 0$  the further restrictions take place $f_{{\mathrm{IR}}} < f_\pm$, and similarly for  $B < 0$ we find that $f_\pm <  f_{{\mathrm{IR}}}$.
Notice that for $B>0$ the proper existence of $\s_{top}^+$ and/or $\s_{top}^-$  determines the following order of the vacua of our quartic $W$: $\s_{{\mathrm{IR}}} < \s_{{\mathrm{UV}}} < \s_{top}^\pm$, which makes possible the construction of a domain wall connecting the two physical vacua $f_\ir$ and $f_\uv$ ($f_\ir > f_\uv$) that are now belonging to the \emph{same} GB-like physical region. Instead, in the case of $B<0$ they are both placed into the same maximal scale physical region (and now $f_\ir > f_\uv$).

\subsection{Properties of DWs solutions}
The easiest way to solve the DW's $I^{st}$ order equations consists in first integrating the following  ``ratio" of eqs. (\ref{sys}):
\begin{equation}
\frac{d \s^2}{d A} = - \frac{16 B^4 L_+^2 L_-^2}{(d-2)^3} \; \frac{\s^2(\s^2 - x_0)}{(\s^2 - x_0)^2 + D} \; \prod_{i = 1}^8 ( \s^2 - \s^2_i ) \; , 
\end{equation}
i.e. to consider the matter field $\s=\s(A)$ as a function of $A$. We have introduced the parameters $\s_i^2 \equiv u_i + x_0  $  with    $  u_1 = - u_2 = u_+ \; ; \;\; u_3 = - u _4 = \tilde{u}_+ \; ; \;\; u_5 = - u_6 = u_-$ and $u_7 = - u_8 = \tilde{u}_- $   given by : 
\begin{eqnarray}
u_\pm = \sqrt{ \frac{(d-2)}{B L_\pm} - D } \; ; \;\; \tilde{u}_\pm =  i  \sqrt{ \frac{(d-2)}{B L_\pm} + D} , \nonumber
\end{eqnarray}
that are related to the positions of the "topological vacua", i.e. all the (real or/and complex numbers, depending on the values of $\la$ and $\m$) algebraic solutions of eqs. (\ref{sigma top eq}) above. The result of the integration
\begin{equation}
e^{A(\s)} = e^{A_\infty} \, |\s |^{- 1/s_{{\mathrm{IR}}}} \; \mid \! \s^2 - x_0 \! \mid^{- 1/s_{{\mathrm{UV}}}} \; \prod_{i = 1}^{8}  \mid \! \s^2 - \s_i^2 \! \mid^{- 1 / s_{top}^i} \; ,   \label{Solution} \\
\end{equation}
provides a rather compact and suggestive form for the scale factor of the DW metrics (\ref{dw}) as a function of the matter field, with the following explicit values of the ``critical exponents" $s_A$ (\ref{criti}):
\begin{eqnarray}
s_{{\mathrm{UV}}}& =& 16 B x_0 L_{{\mathrm{UV}}} \left( 1 - \frac{L_+^2} { L_{{\mathrm{UV}}}^2} \right) \left( 1 - \frac{L_-^2 }{ L_{{\mathrm{UV}}}^2}\right )   ; \label{s uv} \\
 s_{{\mathrm{IR}}} &=& - 8 B x_0 L_{{\mathrm{IR}}} \left( 1 - \frac{L_+^2}{  L_{{\mathrm{IR}}}^2} \right) \left( 1 - \frac{L_-^2 }{ L_{{\mathrm{IR}}}^2}\right) ; \label{s 0}   \\
 s_{top}^p &=& - 64 (d-2)^{-1}  B^2 x_0^{-2} L_+^2 (\s^2_p - x_0)^2 \, \s^2_p \, \left( 1 - \frac{L_-^2}{  L_+^2} \right) ; \; p = 1, \dots , 4; \label{s top +} \\
 s_{top}^q &=& - 64 (d-2)^{-1}  B^2 x_0^{-2} L_-^2 (\s^2_q - x_0)^2 \, \s^2_q  \, \left( 1 - \frac{L_+^2}{ L_-^2} \right) ; \; q = 5, \dots , 8, \label{s top -}  
\end{eqnarray}
According to their definition, which we have introduced in Sect.3 for generic form of the superpotential, they are related to the effective vacua $\s_k$ masses $m^2(\s_k)$ and thus  to the scaling dimensions $\Delta_k$ of certain ``dual" conformal fields from the conjectured  $CFT_{d-1}$'s holographically ``attached" to each one of the Lovelock vacua. 

We next integrate the first equation in (\ref{sys}) in order to find $y = y(\s)$. The integral is similar to the one above, giving
\begin{equation}
e^{y} = e^{y_{n.s.}} \, |\s |^{-L_\ir /s_\ir} \mid \! \s^2 - x_0 \!\mid^{-L_\uv / s_\uv} \prod_{i=1}^8 \mid\! \s^2 - \s_i^2 \!\mid^{-L_i / s^i_{top} } \; , \label{e of y}
\end{equation}
where $y_{n.s.}$ is an integration  constant and $L_i = L_+$ for $i = 1, \dots, 4$, $L_i = L_-$ for $i = 5, \dots, 8$. Together with eq. (\ref{Solution}) it provides  an implicit form for the scale factor $e^{A(y)}$ for a family of Lovelock DWs  in the particular case of our quartic superpotential.  Only for very special rational values of $s_k$ the eq. (\ref{e of y}) becomes an algebraic polynomial equation in $\s$ (with coefficients depending on y), whose roots give the explicit form of $\s = \s(y)$. In this case, by simple substitution in (\ref{Solution}) of these $\s(y)$'s one can derive the explicit form of $e^{A(y)}$ as well. 

It is very important to comment here the fact that the above \emph{relatively implicit} form (\ref{Solution}), (\ref{e of y}) of the DWs metrics reflects the specific (but quite natural) \emph{gauge fixing} used in the particular form of our DWs anzatz (\ref{dw}), i.e. in choosing the ``lapse" factor as $g_{yy}=1$. What is essential in the definition of the flat DWs (independently on the form of the gravitational action) is that it must have $SO(d-2,1) \rtimes T_{d-1}$ symmetry, representing the Poincare group in $(d-1)-$dimensions. Therefore we can chose as a ``radial" coordinate an arbitrary function g(y). There exist, however, one very special (and extremely important for the $AdS/CFT$ and the RG holographic flows applications) \emph{choice},  namely to take the scalar field $\sigma(y)$ as a new coordinate:
\begin{eqnarray}
ds^2&=&\frac{\kappa^2d\sigma^2}{4 (W')^2 \C(W)^2}+e^{2A(\sigma)}\eta_{ij}dx^idx^j,\label{scoord}\\
A(\sigma)&=&-\frac{\kappa^2}{2(d-2)}\int^{\sigma}d\eta\frac{W(\eta)}{W'(\eta)C_0(\eta)}.\nonumber
\end{eqnarray}

We have used eqs.(\ref{sys}) in order to realize the change of the ``radial" variables  $y\rightarrow \sigma$ from eq.(\ref{dw}) to eq.(\ref{scoord}).
Remembering that the DWs matter field $\s$ represents the ``running" (with the energy scale) coupling constant in the dual $QFT_{d-1}$ \cite{VVB}, \cite{rg}, it is not difficult to recognize that the "implicit" form of our DWs solution (\ref{Solution}) is in fact related (proportional) to the ``correlation length"\footnote{ see for example refs. \cite{nmg} and \cite{holo} where such interpretation of the DWs scale factor in $d=3$ New Massive Gravity models as the inverse of the singular part of the reduced free energy of the ``holographic duals" two-dimensional statistical mechanics models is  justified} in the Wilson RG description of the corresponding dual $QFT_{d-1}$ models 
\cite{CS-new}. It is worthwhile to stress here that the practical realization  of such choice of the $QFT_{d-1}$ coupling constant as radial (extra) coordinate, as well as the further  explicit ``reconstruction" of the (non-perturbative) beta-function of this $QFT_{d-1}$ \cite{rg}, \cite{2} is indeed impossible without the first order eqs.(\ref{sys}) and of the superpotential $W(\s)$.

 Notice that all the information about the  dependence of the above DW solutions on the Lovelock couplings ($\la$ and $\m$) is hidden in the explicit form of  $\s_i^2$ and in the effective topological scales $L_{\pm}$ as one can see from the relation, say $L_+^2 = (1 + \sqrt{1 - 3 \la}) L_{0+}^2/ 3$ for the case of  $\m_+$ model, independently of whether or not we have topological vacua at the considered region, i.e. when one or both $L_{\pm}$ are purely imaginary.

 An important property of the DW solutions (\ref{Solution}) and (\ref{e of y}) is that they contain \emph{all the boundary data} of the considered gravity-matter model. By construction the above solutions include all the information about the complete set of vacua $(\sigma_k,L_k,s_k)$ (those with $\s<0$ as well) that are representing boundaries or horizons  of the corresponding $(a)AdS_d$ space-times depending on the signs of $s_k$, as one can see by taking the ``near-critical" limits of the scale factor $e^{2A(\s)} \sim ( \s - \s_k)^{-\frac{2}{s_k}}$. The horizons corresponds to $s_k < 0$ due to the fact that in this case $e^{2A} \rightarrow 0$ is not divergent, while for $s_k >0$ we have  $e^{2A} \rightarrow \infty$ and therefore such vacuum represent  an $AdS_d$  boundary. Thus we can organize all the vacua positions $\s_k$  into a set of consecutive intervals $(\s_l , \s_{l+1})$, $l =1,2,\ldots$ 
and then each one of these intervals is corresponding to one particular \textit{different} DW. Hence the complete set of DW's solutions exhausting all the admissible boundary conditions form, by construction, a finite ``chain'' of consecutive DW's with common boundaries and/or horizons. The individual members DW$_{l,l+1}$ are selected by the ``initial" (in $y$) values $\s_0=\s(0)$ (necessary to define one solution of eq.(\ref{e of y})) depending on which of the intervals it is belonging, i.e. $\s_0 \in (\s_l , \s_{l+1})$. Another remarkable property of the above DW's solution concerns its asymptotic behavior at $ \s \rightarrow \pm \infty$, namely $e^{2A} \rightarrow e^{2A_{\infty}}=const.$ as a consequence of the even more remarkable fact: $s_{{\mathrm{IR}}}^{-1} + 2 s_{{\mathrm{UV}}} ^{-1} + 2 \sum_i (s_{top}^i)^{-1} = 0$. 
Notice that, due to the reflection symmetry of quartic superpotential, all the vacua (excepts $\s_{IR}=0$) are ``doubled", i.e $\pm \s_k$ and $\pm \s_{UV}$ have equal critical exponents, which is explaining the presence of the extra factor 2 in a part of them. As one can easily  verify from eq. (\ref{curvature}), this limit corresponds to an infinite value of the curvature (although the scale factor is finite) representing a naked singularity (n.s.) at $y_{n.s.}$ (as it follows from eq.(\ref{e of y})). Such ``singular domain walls" corresponding to ``initial" condition $\s_0 \in (\s_l , \infty)$ are in fact present in very few of the cases analyzed in Sect.3.: those of $\la$ and $\m$ both negative (of no topological vacua at all) and under certain extra conditions in few  cases corresponding to regions of ``maximal" scale. Their importance in the description of certain massive phases (and massive directions of the RG flows) in the $QFT_{d-1}$ dual to  the gravity-matter models in the case of $d=3$ New Massive Gravity was first demonstrated in refs. \cite{nmg} and \cite{holo}. 
 
The above discussion together with the classification of  all the vacua presented in Sect.3. suggests that for different shapes of the superpotential and for different values of $\la$ and $\m$ we can have distinct sequences of DW's corresponding to qualitatively different $(a)AdS$ geometries: the \emph{stable} physical ones $AdS_d(IR)/AdS_d(UV)$; the \emph{unstable} ones relating one physical to one topological vacua or those between two topological vacua and finally  the \emph{singular} ones as for example $AdS_d(UV)/n.s.$. How many different DW's we can have in the different regions (see Figs.1) indeed depends on the superpotential. The answer is always encoded in the analytic properties (cuts, poles, zeros, etc.) of the DW scale factor, determined by the signs and the values (real or imaginary) of the critical exponents as well as on the number (and the positions) of these singularities, related to the real or/and complex nature of the topological scales $L_\pm$.
Take for example  $\m_+ > 0$ and $\la < 1/4$ where we have only one topological vacua since  the scale $L_-$ is imaginary. In this case  the product in (\ref{Solution}) for $i = 5, \dots, 8$ is non-singular and the corresponding DW's solution describes two DW's (of common boundary):  one ``physical" and one ``topological'' DW's (relating $\s_{UV}$ to the topological vacuum $\s_{top}^+$  of minimal scale $L_+$) both belonging to the GB-like region. Consider next the case  $1/4 < \la < 1/3$ and  $\m_+ < 0$ when  the both  scales $L_\pm$ are real and therefore we have  four singularities\footnote{Notice that from the eight ``topological vacua" positions $\s^2_i$, only four are real and represent the true vacua solutions. Nevertheless, the four complex  $\s^2_i$ are not arbitrary but complex conjugate each to other and still lead to real expressions in eq. (\ref{Solution}) due to the fact that  the corresponding ``critical exponents"  are  imaginary numbers. The same is true  when one of  the scales $L_\pm$ is is imaginary.}
under the product symbol in (\ref{Solution}). Now we have few options of DW's connecting physical vacua to topological vacua with either minimal or maximal scales. Thus, in agreement  with the conclusions of Section 3, we can have one GB-like minimal scale region, and one maximal scale region. The same  is true for the $\m_-$ model, for all $\la > 0$ where both $L_\pm$ are real. Finally, for $\la<0$ and $\m_- < 0$, both scales $L_\pm$ are imaginary, hence the whole part of the solution under the product symbol is nonsingular, i.e. there is no topological vacuum at all.

The  next question is about the different types of boundary conditions, and hence concerns the particular form  of the ``DW's chains'' admissible in each of the two types of physical regions, described above. First, we notice that independently on the region, the existence of $f_\ir$ and $f_\uv$ allows us to construct DWs interpolating between two AdS$_d$ vacua with different radii $L_\ir$ and $L_\uv$ 
and with critical exponents $s_{{\mathrm{IR}}}$ and $s_{{\mathrm{UV}}}$ of \emph{different} signs. Therefore the sequence of DW's leads to the follwing  ``chain'': AdS$_d$($\s_\ir$)/AdS$_d$($\s_\uv$)/AdS$_d$($\s_{top})$ for the GB-like case, while  in the maximal scale region the sequence is with  $AdS_d(\s_{top})$ and $AdS_d(\s_{IR})$  of inverted  positions. In the case  when we have  no one topological vacua, the b.c.'s permit the presence of singular DW's, i.e. two DW's of common boundary: $AdS_d(IR)/AdS_d(UV)/n.s.$ that are relevant for describing second order phase transitions form massless to massive phase in the conjectured dual $QFT_{d-1}$, as we have mentioned above.

\subsection{Marginally  degenerated DWs}
 The case when one of the $s_k$ is \emph{vanishing} requires special attention. As we have shown at the end of Sect.3., it takes place when the corresponding extremum $\s_k$ of $W$ is second (or higher) order zero of $W'(\s)$, or in the case when one of the simple $W$ extrema is coinciding with one or more of the topological vacua either when few of the topological vacua are colliding. Then an important qualitative change in the asymptotic behavior of the scalar field and of the scale factor do occur. In order to demonstrate the nature of this phenomena, without introducing any essential changes in the structure of quartic superpotential (as for example when $x_0=0$, i.e $\s_{UV}=\s_{IR}$) we consider few particular ``critical" points in the parameter space: (\emph{i}) $L_+=L_-=L_{CS}$ of two coinciding topological vacua\footnote{ As it well known this special case of the extended cubic Lovelock gravity for $d=7$ can be identified with the well known AdS-Chern-Simons \cite{espanha},\cite{zanelli} action (coupled to scalar matter) of fundamental scale $L^2_{CS}$ }, i.e. for $\la=1/3$ and $\m=-1/27$, and (\emph{ii}) the case $L_{UV}=L_+$ as well. 
 
 Consider first the case of two coincident topological vacua. The form of the scale factor can be obtained  by taking the limit $\s_p \rightarrow \s_q$, with $p = 1, \dots , 4$ and $q = p + 4$. It can be seen from (\ref{s top +}), (\ref{s top -}) that $s_p , s_q \rightarrow 0$ and $s_p / s_q < 0$,  
thus in the vicinity of the vacua one can make the parametrization $s_p = - s_q = \eta \, \e$ and $\s^2_q - \s^2_p = \rho \e$, being $\eta$ and $\rho$ independent parameters. It follows that, when $\e \rightarrow 0$, 
\begin{equation}
(\s^2 - \s^2_p)^{- \frac{1}{s_p}}(\s^2 - \s^2_q)^{- \frac{1}{s_q}} \rightarrow e^{- \frac{\rho}{\eta (\s^2 - \s^2_p)}} \; .
\end{equation}
For $p = 1$, one has, with $\e = L_+^2 - L_-^2$,
$$\eta = - \frac{64 B^2}{(d-2)} \s_1^2 (\s_1^2 - x_0)^2 \; , \;\; \rho = (d-2) \left\{ 4 B L_{CS}^3 \sqrt{ (d-2)/ B L_{CS} - D} \right\}^{-1} , $$
and similarly for the other cases. Thus, we find  that scale factor has the following  form:
\begin{eqnarray}
&&e^{A(\sigma)} \sim |\sigma |^{-1/ s_{IR}}  |\sigma^2 - x_0|^{-1/s_{UV}}  \exp \left[ \frac{(d-2)^2}{254B^3L_{CS}^3 \, \s^2_1(\s^2_1 - x_0)^2 \sqrt{\frac{(d-2)}{BL_{CS}} - D}} \; \frac{1}{\sigma^2 - \s^2_1}\right] \times   \nonumber\\
&&\times \exp\left[ \frac{-(d-2)^2}{254B^3L_{CS}^3 \; \s_2^2(\s_2^2 - x_0)^2\sqrt{\frac{(d-2)}{BL_{CS}}-D}}\; \frac{1}{\sigma^2- \s^2_2}\right] \times   \nonumber\\
&&\times \exp \left[ \frac{L_\ir L_{UV} (2x_0-\sigma^2)}{127(d-2)\left(L_{UV} + L_{CS} \right)\left(L_{IR}+ L_{CS}\right) \left( 1+\frac{BL_{CS}}{(d-2)}\left[(\sigma^2-x_0)^2+D\right]\right)}\right], \label{topdege}
\end{eqnarray}
where we have written the topological vacua contributions explicitly as real numbers. Since $s_k = 0$, the characterization of the vacuum as describing $AdS_d$ boundaries  or/and  horizons depends on whether $\s$ approaches $\s_k$ from the right or from the left. For $\sigma_2$ it represents  a horizon when is approaching from the right and a boundary if it approaches from the left, thus  preserving  the nature of the vacuum before the limit has been taken. Notice that  we \emph{can not} interpret these two DW's as forming a ``chain" any more. 
Together with the changes in the scalar field asymptotic properties, the scale factor is representing now an essential singularity. It is expected that in the dual $QFT_{d-1}$ these marginal critical points are describing certain infinite order phase transitions, similarly to the $d=3$ case based on the New Massive Gravity (see ref.\cite{holo}). Hence in the limit of two coinciding topological vacua and similarly in the cases when one physical and one or two topological vacua are colliding, the essential singularity is replacing two or more brunch-cuts  of the scale factor, describing eventually the transformation of few second order phase transition points into one of infinite order.

\begin{figure}[ht]
\centering
\subfigure[]{
\includegraphics[scale=0.6]{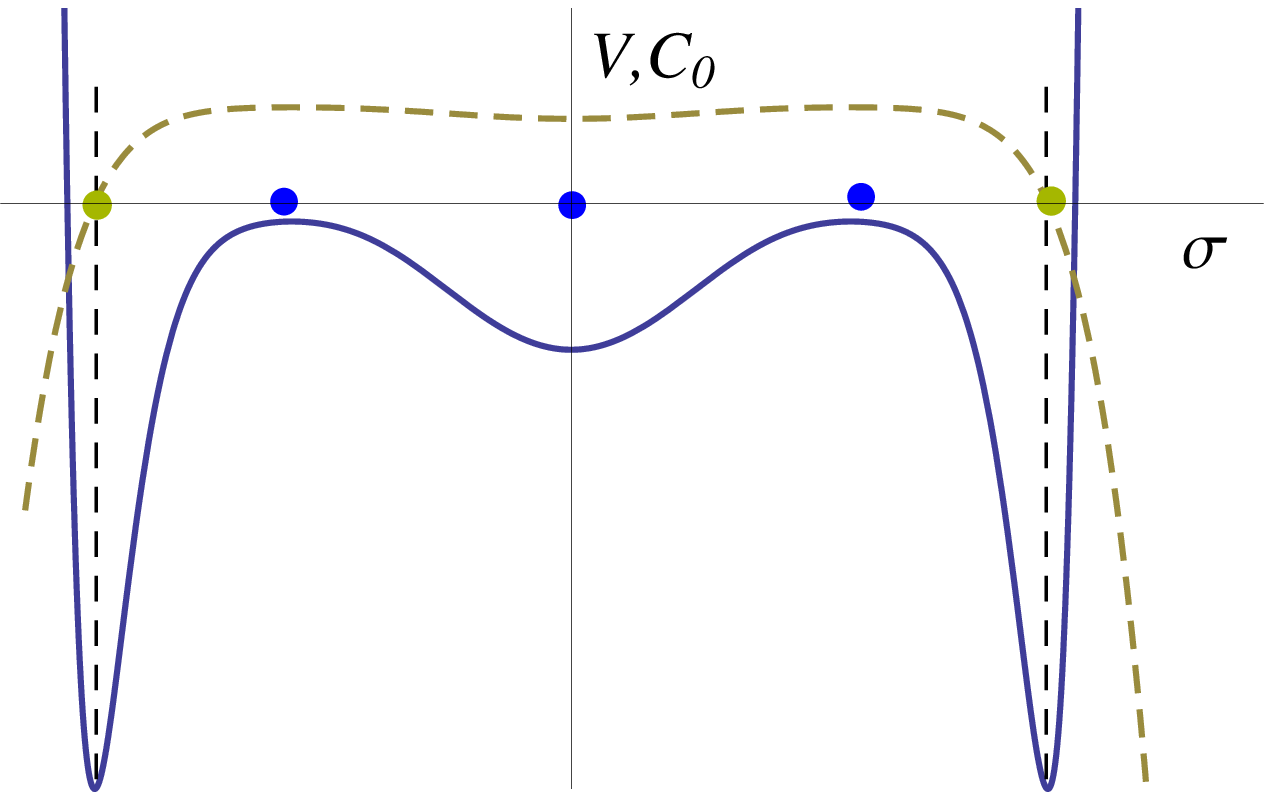}
\label{fig:2a}
}
\subfigure[]{
\includegraphics[scale=0.6]{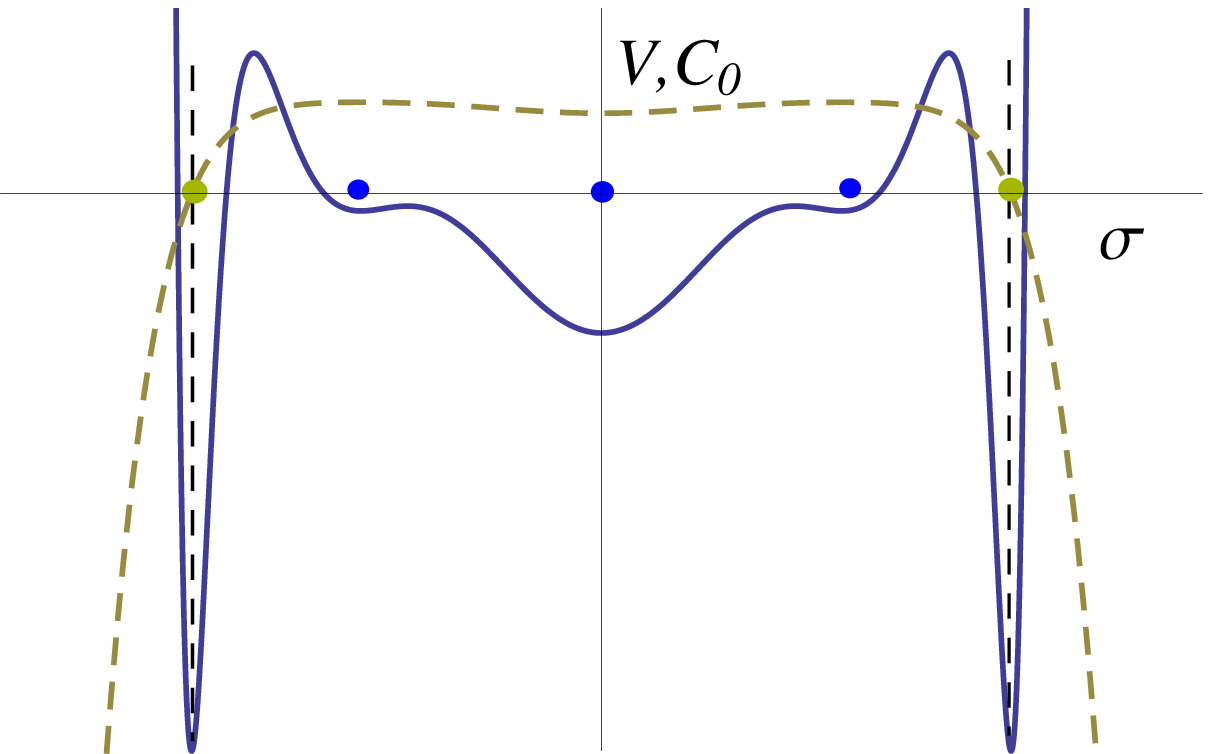}
\label{fig:2b}
}
\label{fig:2}
\caption{The matter potentials $V(\sigma)$ in blue for quartic superpotentials and $C_0$ are the dashed yellow curves: $(a)$ $0<s_{UV}<d-1$; $(b)$ $s_{UV}>d-1$. The yellow dots mark the zeros of $C_0$ representing topological vacua.}
\end{figure}


Another interesting degenerated case is given by the limit when one physical vacua is approaching the topological one: $L_{UV} = L_+ \equiv L_*$. Since $L_{UV} = (d-2)/BD$, this is equivalent to $\s^2_1 = \s_2^2 = x_0$, and therefore it represents  the limit between one physical and two topological vacua. Defining $\e = (L_{UV}/L_+) - 1$, one finds $\s_1^2 = x_0 + \sqrt{D \e}$, $\s_2^2 = x_0 - \sqrt{D \e}$, $s_{UV} = 32 x_0 B L_* (1 - L_-^2 / L_*^2) \e$, $s_{top} = - 2 s_{UV}$ and also that:
\begin{eqnarray}
&&(\s^2 - x_0)^{-1/s_{UV}} \left[ \s^2 - (x_0 + \sqrt{D\e}) \right]^{1/s_{top}^1} \left[\s^2 - (x_0 - \sqrt{D\e}) \right]^{1/s_{top}^2}  \sim \exp \left[\zeta_* \frac{1}{(\sigma^2-x_0)^2} \right], \nonumber
\end{eqnarray}
where $ \zeta_* = -L_*D[32 x_0 B (L_*^2 -L_-^2)]^{-1}$.
Thus, the final  form of the scale factor in this case is given by:
\begin{equation}
e^{A(\sigma)} \sim |\sigma |^{1/s_{IR}}  \exp \left[ -\frac{L_* D}{32 x_0 B \left(L_*^2 - L_-^2 \right) \left( \sigma^2 - x_0 \right)^2} \right] \prod_{i=3}^8 |\sigma^2 - \s_i^2|^{-1/s_{top}^i}. 
\end{equation}
Notice that at this particular doubly degenerated $L_{UV}=L_+$ point in the parameters space, the essential singularity  $exp(2A)\approx exp(\frac{2\zeta_*}{(\s^2-x_0)^2})$ turns out to be ``stronger"
compared to the simpler degenerated case  described by eq. (\ref{topdege}).

\subsection{BF restrictions on Gauss-Bonnet DWs}

In order to analyse the effect of the BF-unitarity restrictions (\ref{news}) on the shape of the superpotential (and of the $V(\s)$ as well) we consider the particular limiting case  $\la = 1/4$, $\m_+(\la) = 0$ corresponding to GB gravity coupled to scalar matter with fundamental scale $L^0_{top}=\frac{L_{top}}{\sqrt{2}}$. As we have shown in Sect.3. we have only one topological vacua (and no more  $L_-$ scale is present), hence we have less roots of the topological vacua equation (\ref{sigma top eq}), namely  $\s_i$, $i = 1, \dots, 4$. The solution for the scale factor of this  GB gravity coupled to matter can be found by taking the $\m \rightarrow 0$ limit of the Lovelock DW's scale factor (\ref{Solution}):
\begin{eqnarray}
&&e^{A(\s)} = e^{A_\infty} \s^{- 1/s_{{\mathrm{IR}}}} \; \mid \! \s^2 - x_0 \! \mid^{- 1/s_{{\mathrm{UV}}}} \; \prod_{i = 1}^{4}  \mid \! \s^2 - \s_i^2 \! \mid^{- 1 / s_{top}^i} \; , \\
&& s_{{\mathrm{UV}}} = 16 B x_0 L_{{\mathrm{UV}}} ( 1 - L_{top}^2 / L_{{\mathrm{UV}}}^2 ) ; \;\; s_{{\mathrm{IR}}} = - 8 B x_0 L_{{\mathrm{IR}}} ( 1 - L_{top}^2 / L_{{\mathrm{IR}}}^2 ) ; \\
&& s_{top}^i = - 64 (d-2)^{-1}  B^2 a^{-4} L_{top}^2 (\s^2_i - x_0)^2 \, \s^2_i \,; \; i = 1, \dots , 4;\label{gb} 
\end{eqnarray}
We next consider the simplest ``physical" case defined by the following restrictions on the scales:
\begin{eqnarray}
L^2_{UV}>L^2_{IR}>L^2_{top} \label{gbfis}
\end{eqnarray}
which indeed describes (for $\s>0$) a ``chain" of two different GB domain walls of common boundary: the physical one  $AdS_d(IR)/AdS_d(UV)$ 
and  ``phys-top" one  $AdS_d(UV)/AdS_d(top)$ which is relating the physical UV vacua to the topological one $\s_{top}$. Given the explicit form of all the critical exponents in terms of the superpotential parameters as in eqs.(\ref{gb}) above, we note that when the condition (\ref{gbfis}) is satisfied we always have that $s_{UV}$ is positive and both $s_{IR}$ and $s_{top}$ are negative. We further investigate  the difference between the two distinct cases corresponding to: (a) $0<s_{UV}<d-1$, i.e. the ``stronger" BF condition $-\frac{(d-1)^2}{4 L_{UV}^2}\le m_{\sigma}^2(\sigma_{UV}^*)<0$ and (b) $s_{UV}>d-1$, i.e. when $ m_{\sigma}^2(\sigma_{UV}^*)>0$  discussed  at the end of  Sect.3. Due to the difference between the extra BF-like restrictions imposed on  the superpotential parameters, these two cases provide GB domain walls of two completely different matter potentials $V_{(a)}(\s)$ and $V_{(b)}(\s)$ (as shown on Figs. \ref{fig:2a} and \ref{fig:2b}), although the inverted ``double-well" form of the superpotentials $W_{(a)}$ and $W_{(b)}$ is indeed preserved. One can easily understand the presence  of two more ``maxima"  in the potential $V_{(b)}$  originated by the requirement that  now $m^2_{UV}(b)>0$ (differently from the $V_{a}$ case) and also that the $m^2_{IR}$ and $m^2_{top}$ are  positive (as always). Hence the new maxima must appear in between the ``neighbours" minima $\s_{UV}$ and $\s_{top}$ (in the left) and $\s_{IR}$ (in the right). In this (b) case the GB domain walls are interpolating between two minima of the matter potential (similarly the the ``kinks" in the classical field theories on flat background), while in the case (a) they are relating one  maximum with the neighbouring minima. Let us mention that the case (a) is indeed the one whose DW's are used to describe the ``holographic" RG flows and the phase transitions in $(d-1)$ dimensional $QFT_{d-1}$ dual to the Gauss-Bonnet Gravity interacting with scalar matter of superpotential $W(\s)$. The case (b) as we have mentioned in Sect.3 does not lead nor to unitary $QFT$'s neither is giving rise to \emph{relevant } operators necessary in order to have RG flows. We are also plotting together with the potential the values of the $\C(\s)$-function of the GB domain walls in order to demonstrate its main property (known as c-theorem): it indeed is decreasing  from UV- to the IR-vacuum  and also in the other direction towards the to topological vacuum where it becomes zero\footnote {Similar observations, arguments and conclusions concerning the particular properties of the GB DW's and also about their use in the description of certain  RG flows are presented in Apps. B and C of the recent paper by Myers and Singh \cite{Myers-new} we have received few days before the completion of this  paper. Some partial results where presented by GMS in his talk at the workshop ``Quantum Field Theory and Quantum Gravity" 02/13-15/2012 held in Ubu, Brazil \cite{ubu}.}. The same phenomena take place in the case of the Lovelock gravity coupled to matter. We leave the complete description of the consequences of the ``stronger" BF condition in this case to our forthcoming paper devoted  to the RG flows in  the $QFT$'s duals to  Quasi-Topological Gravity coupled to scalar matter.

\section{Discussion}

\setcounter{equation}{0}

Using the effective Lagrangian method, we have  obtained the  second order field equations for DWs of the extended cubic Lovelock-matter gravity action in $d$ dimensions. The way we have derived the corresponding  $I^{st}$ order eqs. (\ref{sys}) by introducing an appropriate superpotential $W$, suggests that our cubic Lovelock results can be easily extended for generic Lovelock gravity coupled to matter, due to the fact that the main ingredient $\C(W)$ is already known for generic extended Lovelock Gravity models \cite{espanha},\cite{2}. It is worthwhile to mention one interesting  problem for further research that concerns the Supergravity origin of the first order system. Namely, its relation to the constant Killing spinor equations in the supersymmetric extensions of these Lovelock-matter gravity models (similarly to the well known case of Einstein-matter supergravity \cite{6},\cite{5}). Therefore what we need is an appropriate supersymmetric version of the  Quasi-topological Gravity with chiral matter supermultiplets added, that is expected to reproduce  the superpotential, its specific  relation to the matter potential and the  $I^{st}$ order equations (\ref{sys}) as well.
 
An important consequence of the explicit form of the first order equation for the matter field $\s$ involving $\C(W)$  is that one can always find an appropriated range of values of the gravitational couplings $\la$ and $\mu$ (not both negative), such that  at least one of the vacua of the considered  gravity-matter model (\ref{qtop}) is of topological nature, i.e. with $C_0 = $0. As we have shown in Sect.3.2, the proper existence of topological vacua  introduces natural smallest or largest AdS$_d$ scale(s) that can be chosen as a fundamental scale $L^2=L^2_{0,top}$ present in the action (\ref{qtop}). It also determines two particular families of models  corresponding to specific relations between the  Lovelock couplings $\lambda$ and $\mu$.
Further restrictions on the gravitational couplings and, by consequence, on the shapes of $V$ and $W$, are found by imposing the stability (causality) $\C>0$ and BF-unitarity conditions (\ref{news}) that selects few ``domains" of  physical Lovelock-matter vacua.   

The ``chains" of physical and topological explicit DWs solutions, presented in Sect.4 for the case of quartic ``double-well" superpotential, that interpolate between different extrema of the matter potential, are known to be the main tool in the investigation of the Renormalization Group flows \cite{VVB},\cite{rg} and of the phase transitions that take place in the conjectured holographic dual $QFT_{d-1}$ \cite{2},\cite{Myers-new}.
As we have mentioned in Sect.4.2, the relatively implicit form (\ref{Solution}) of the Lovelock DWs scale factor  as a function of the scalar field $\s$ represents one of the most important ingredients for the further analysis of the nature of the critical phenomena that occur in the corresponding $QFT_{d-1}$. The rather ``natural" choice of the $QFT$'s coupling constant as the (extra) radial coordinate
\begin{eqnarray}
ds^2=\frac{\kappa^2d\sigma^2}{4 (W')^2 \C(W)^2}+e^{2A(\sigma)}\eta_{ij}dx^idx^j,\quad \quad e^{A(\s)}\sim |\s|^{-1/s_{IR}} \mid \! \s^2 - x_0 \! \mid^{-1/s_{UV}}  \prod_{i = 1}^{8}  \mid \! \s^2 - \s_i^2 \! \mid^{-1/s_{top}^i}\nonumber
\end{eqnarray}
of the dual $(a)AdS_d$ space-time (i.e.the Lovelock DWs geometry) provide in fact the ``explicit" forms of all the important for the RG description quantities as for example: the ``correlation length", the singular part of the reduced free energy, the $\beta$-function,  etc.\footnote{see the J.Cardy textbook \cite{cardy} for the definitions and for the rather simple and illuminating introduction in the Wilson RG and its applications to QFT and statistical mechanics problems}. We have to remind  once more the important role played by the first order eqs. (\ref{sys}) and of the superpotential $W(\s)$ in the practical realization of the ``off-critical" holography \cite{VVB},\cite{rg}, i.e. for the extracting of the off-critical properties of the dual $QFT$ from the corresponding  Domain Walls geometric data. All these results are particularly relevant for $d = 5$. In this case our Lovelock DWs  constructions provide together with the explicit form of the beta-function, the two different \emph{conformal anomalies functions} $c(\s)\ne a(\s)$ as well \cite{2},\cite{Myers-new}. Thus, the $d=5$ cubic Lovelock DWs allows us to establish the conditions for the validity of the a/c-theorems in the case when both vacua (critical points) are physical, i.e. for their decreasing from the UV to IR scales (see for example Fig.2 for GB case). The complete discussion of the RG flows and a/c-theorems in the $QFT_4$ duals to $d=5$ cubic Quasi-Topological Gravity coupled to scalar matter will be presented in our forthcoming paper \cite{CS-new}.

The methods we have used in the  constructions of the Lovelock  DWs  in the particular examples of quadratic and quartic superpotentials are quite general. The same methods are perfectly working in the case of more complicated (non-polynomial) forms of the  superpotential \cite{vicosa},\cite{rg} as well as in the cases involving  more then one scalar matter fields. For example, the flat Domain Walls of the particular ``stringy induced" superpotential $W(\s) = B \cosh (\kappa\s) [ 2\delta - \cosh(\kappa\s)]$, turns out to share many of the properties of those of  the ``double-well" superpotential described in Sect.4. It is also worthwhile to notice the relevance of the \emph{quadratic} superpotential Lovelock DWs we have studied in App.B, reminding the well known and largely explored fact: namely,  that an arbitrary superpotential (that has at least one extrema) can be approximated near to the boundary or/and horizon by such quadratic superpotential. Therefore our results provide the universal near-boundary form of the cubic Lovelock-matter model Domain walls for arbitrary superpotentials.

Our final comment is about the special $d = 4$ case, where no consistent analogue of the Quasi-Topological gravity is known. As we have shown in Sect.2, the ``reduced"  cubic Lovelock action (\ref{gbl-action}) with particular choice  (\ref{valores}) of the parameters: $\gamma_4=1/3$ and $\alpha_4=-36/7=-\beta_4$  leads to second order DWs equations too. Evidently, all our discussions and DWs constructions take place in this $d=4$ case as well. Their particular importance is however in the cubic Lovelock extensions of the cosmological models. As it well known  any given  $AdS_4$ DWs solution can be easily transformed into  $dS_4$ FRW cosmological solution by  ``analytic" continuation. Thus, the standard physical DWs are giving rise to interesting bounce-like solutions, while the singular ones - to big-bang or big-crunch FRW solutions. It turns out that the cubic Lovelock gravitational interaction terms (for appropriate values of $\m$) are responsible for the few acceleration periods \cite{LoveCosmo} that these Lovelock-FRW solutions do represent, without  adding of any fluids with strange (ghost-like) stress tensors. 

\newpage

\appendix

\noindent{\Large {\bf Appendices}}

\vspace{0.5cm}

\section{The effective Lagrangian Method}\label{apexA}
\setcounter{equation}{0}
There are two distinct ways of deriving the equations of motion (\ref{eqmat}), (\ref{eq2}) and (\ref{constr}). The standard one consists in the variation of the action (\ref{gbl-action}) with respect to the metric (and w.r.t $\s$ as well) and further by  substituting the DWs anzatz (\ref{dw}) in these ``higher" order differential equations. The second method, we have used, is known under the name ``effective Lagrangian'' method \cite{town1}.

We begin with the following modified (i.e. non-completely gauge fixed) ansatz (\ref{met}) for the metrics:
$$ds^2 = f^2(y) e^{2 (d-1) A(y)} dy^2 + e^{2A(y)} \eta_{ij} dx^i dx^j ,$$
with the presence of the arbitrary function $f(y)$, thus leaving the  ``laps" factor  $g_{yy}$ undetermined. Using this metric, we shall calculate all the curvature terms in the action (\ref{gbl-action}). We find the following non-vanishing connections:  
$$ \Gamma_{yj}^i = \dot{A} \delta_j^i ; \; \Gamma_{ij}^y = - \frac{\dot{A}}{f^2} e^{-2(d-2)A} \, \eta_{ij} ; \; \Gamma_{yy}^y = \frac{\dot{f}}{f} + (d-1) \dot{A} , $$ 
and further we substitute them in the expression for  the curvature tensors components. After some lengthy calculations, we obtain
\begin{eqnarray}
R_{yy}&=&(d-1)\left(-\ddot{A}+\frac{\dot{f}}{f}\dot{A}+(d-2)\dot{A}^2\right), \ \ R_{ij}=-\frac{1}{f^2}e^{-2(d-2)A}\left(\ddot{A}-\frac{\dot{f}}{f}\dot{A}\right)\eta_{ij},\nonumber\\
R&=&g^{\mu\nu}R_{\mu\nu}=2(d-1)(d-2)\frac{e^{-2(d-1)A}}{f}\left(\frac{1}{2}\frac{\dot{A}^2}{f}-\frac{1}{(d-2)}\frac{\ddot{A}}{f}+\frac{1}{(d-2)}\frac{\dot{f}}{f}\frac{\dot{A}}{f}\right)\nonumber\\
&=&2(d-1)(d-2)\frac{e^{-2(d-1)A}}{f}\left(\frac{1}{2}\frac{\dot{A}^2}{f}-\frac{1}{(d-2)}\frac{d}{dy}\left(\frac{\dot{A}}{f}\right)\right),\nonumber
\end{eqnarray}
We next calculate the invariants containing quadratic in the curvature terms:
\begin{eqnarray}
&&\sqrt{|g|}\left(R^{\mu\nu}R_{\mu\nu}-\gamma_dR^2\right)= \nonumber \\
&&(d-1)\frac{e^{-2(d-1)A}}{f^3}\left(d-4(d-1)\gamma_d\right)\left[ \ddot{A}^2+\left(\frac{\dot{A}\dot{f}}{f}\right)^2-2\frac{\dot{f}}{f}\dot{A}\ddot{A}\right] +\nonumber\\
&&+(d-1)^2(d-2)^2(1-\gamma_d)\frac{e^{-2(d-1)A}}{f^3}\dot{A}^4 -\frac{2}{3}(d-1) \frac{\dot{A}^4}{f^3} \, e^{-2(d-1)A} -\nonumber\\
&&- \frac{2}{3}(d-1)^2(d-2)(1-2\gamma_d)\, \frac{d}{dy}\left(e^{-2(d-1)A}\frac{\dot{A}^3}{f^3}\right).\nonumber
\end{eqnarray}
The first term would contribute with higher derivatives (higher than second order) to the equations of motion, so it must vanish. This determines the value of $\gamma_d$:
\begin{eqnarray}
\gamma_d=\frac{d}{4(d-1)} .
\end{eqnarray}
With a little more calculations, we succeed to simplify the cubic Lovelock's invariants to the following form:  
\begin{eqnarray}
&&\sqrt{|g|}\left[ R^3+\alpha_d RR^{\mu\nu}R_{\mu\nu}+\beta_d R^{\mu}_{\nu}R^{\nu}_{\rho}R^{\rho}_{\mu} \right]=\nonumber\\
&&\frac{e^{-4(d-1)A}}{f^5}\Bigg\{(d-1)\left(8(d-1)^2+2d(d-1)\alpha_d +(d-1)^2\beta_d+\beta_d\right)\Bigg[-\ddot A^3+\left(\frac{\dot f}{f}\right)^3\dot A^3+3\frac{\dot f}{f}\dot A\ddot A^2\nonumber\\
&&-3\left(\frac{\dot f}{f}\right)^2\dot A^2\ddot A\Bigg]+(d-1)^2(d-2)\big(12(d-1)+(5d-4)\alpha_d+3(d-1)\beta_d\big)\Bigg[ \dot A^2\ddot A^2+\left(\frac{\dot f}{f}\right)^2\dot A^4\nonumber\\
&&-2\frac{\dot f}{f}\dot A^3\ddot A\Bigg]+(d-1)^3(d-2)^3(1+\alpha_d+\beta_d)\dot A^6\nonumber\\
&&-(d-1)^3(d-2)^2\big(6+4\alpha_d+3\beta_d\big) \frac{f^5}{5}\frac{d}{dy}\left(\frac{\dot A^5}{f^5}\right) \Bigg\}. \nonumber
\end{eqnarray}
It is then clear that the  first two terms would also generate high derivatives in the equations of motion, so their coefficients must vanish as well. This only happens if 
\begin{eqnarray}
\alpha_d=-\frac{12d(d-1)}{d(d+4)-4}, \ \beta_d=\frac{16(d-1)^2}{d(d+4)-4}, \label{valores^3}
\end{eqnarray}
which fixes all the constants in our action (\ref{gbl-action}) to the form (\ref{valores}) as we have declared in Sect.2.

Substituting  these values of $\alpha_d$, $\beta_d$ and $\gamma_d$ in all the curvature invariants we have calculated above, we find the following simple form of our effective Lagrangian (\ref{L-eff}): 
\begin{eqnarray} 
S &=& \int d^{d-1}x \, dy \; \mathcal{L}_{{\mathrm{eff}}} - \nonumber\\
&&-2(d-1)\int \! d^{d-1}x \, dy \; \frac{d}{dy} \Bigg[ \frac{\dot{A}}{f} \left(1+\frac{(d-2)^2 \la_0}{6  m^2} \frac{e^{-2(d-1)A}}{f^2} \dot{A}^2 + \mu_0 \frac{3(d-2)^2}{5m^4}\frac{e^{-4(d-1)A}}{f^4}\dot A^4\right)\Bigg]. \nonumber
\end{eqnarray}
together with  some other terms which can be organized  in  a total derivative and hence they do not contribute to the equations of motion. In order to find the Lovelock DWs  equations we  treat this action as representing a mechanical system, and we further variate it (i.e. we derive the corresponding  Lagrange's equations) with respect to the ``generalized coordinates''\footnote{Notice that in the Lagrangian appear the ``generalized coordinates'' $A$, $\s$ and $f$, as well as their respective ``generalized velocities'' $\dot{A}$ and $\dot{\s}$, but the $\dot{f}$ is not present.} $A$, $\s$ \textit{and} $f$. The arbitrary function $f$ appears as a Lagrange multiplier, and its variation indeed gives the constraint (\ref{constr}):
\begin{eqnarray}
V(\s) = -(d-1)(d-2)\dot{A}^2\left(1+\frac{(d-2)(d-4) \la_0 }{4 m^2}\dot{A}^2+ \frac{(d-2)(d-6) \m_0}{m^4}\dot A^4\right) + \frac{1}{2}\dot{\sigma}^2,\label{ve}
\end{eqnarray}
where  we have finally set $f(y) = e^{-(d-1)A(y)}$. We next deduce the  Lagrange equations for the scale factor $A$ and for the matter field $\s$ which after the same ``gauge" fixing of $f(y)$ take the following simple form: : 
\begin{eqnarray}
&&V(\s) = - (d-2)\ddot{A}\left(1+\frac{(d-2)(d-4) \la_0}{2 m^2}\dot{A}^2+ \frac{3(d-2)(d-6) \m_0}{m^4}\dot A^4\right) \nonumber\\
&&-(d-1)(d-2)\dot{A}^2\left(1+\frac{(d-2)(d-4) \la_0}{4 m^2}\dot{A}^2+ \frac{(d-2)(d-6) \m_0}{m^4}\dot A^4\right),\label{vemat} \\
&&\ddot{\s} + (d-1) \dot{a}\dot{A} = V'(\s) \label{secorder},
\end{eqnarray}
Notice that the final form of the eq. (\ref{eq2}) is then recovered  by substructure of  eq. (\ref{ve}) from the eq. (\ref{vemat}) above.


\section{Lovelock DWs for quadratic Superpotential}\label{apexB}
\setcounter{equation}{0}
The quadratic superpotential provides only one candidate for physical vacuum, so it is impossible to construct a proper \textit{physical} DW solution. It is still possible to construct DWs interpolating between the physical vacuum and a topological one, and such DWs will be the subject of this appendix. However  there is one more important reason to study solutions for this type of superpotential: it is representing in fact the \emph{universal} near-boundary behavior of an arbitrary superpotential in the vicinity of a vacuum (a), i.e. when $W'(\sigma_k^*)=0$ e $W(\sigma_k^*)\neq0$ (in the terminology adopted at Sect. 2):
\begin{eqnarray}
W(\sigma)\approx W(\sigma_k^*)+\frac{W'(\sigma_k^*)}{2}(\sigma-\sigma_k^*)^2.
\end{eqnarray}
In this approximation the scale factor is approaching the proper $AdS$ one, $i.e.$ $A(y)\approx\frac{y}{L}$, and hence we realize  that according to  eq. (\ref{Solution quadratic}) below, the scale factor  takes the following simple form:
\begin{eqnarray}
e^{\frac{y}{L_k}}\approx e^{A_{\infty}}|\sigma-\sigma_k^*|^{-1/s_k},
\end{eqnarray}
which leads us to the specific form of the matter field asymptotic:
\begin{eqnarray}
\sigma\approx \sigma_k^*+ {\mathrm{const.}}\, e^{-s_k\frac{y}{L_k}},
\end{eqnarray}
as declared in Sect.3.(see eq. (\ref{asymp})). In this sense the quadratic superpotential (and its DWs) can represent (in the approximation considered) the most general superpotentials.

We next choose our quadratic superpotential  in the following form:
\begin{equation}
W(\s) = - B \s^2 - D \; ; \;\; D > 0 .
\end{equation}
Again as in the case of quartic superpotential (see Sect.4.) due  of $Z_2$-symmetry, we shall restrict us to consider the case $\s > 0$ only. Analogously to the case of the Higgs-like superpotential, the sign of $D$ is determined by the condition: $W(\s) < 0$ ($\dot{A}>0$) for all $\s$. Then the only physical vacuum  is given by $\s_0 = 0$ and by denoting  $W_0 = W(\s_0)$ we find that
\begin{equation}
W_0 = - D \; ; \;\; f_0 = \frac{L^2 D^2}{(d-2)^2}  \; ; \;\; D = (d-2) \frac{\sqrt{f_0}}{  L} , \label{parameters}
\end{equation}
where $L^2$ is given by the normalization of Sect.3., i.e. $L^2 = L^2_{0+} = \tilde{L}^2_{0-}$.

Depending on the values of $\la$ and $\m$, they might exists  one, two or zero topological vacua (respectively, if $\m > 0$, $\m<0$ and $\la>0$, $\m <0$ and $\la < 0$).

Writing the parameters of $W(\s)$ in terms of $f_0$, we find that  the topological vacua equation takes the form:
\begin{equation}
\s_\pm^2 =  \frac{(d-2)}{BL}\left[ \sqrt{f_\pm} - \sqrt{f_0} \right] \; , \label{sigma top eq quad}
\end{equation}
which leads us to the  conclusion  that: (\textit{i}) if $B > 0$, then $f_0 < f_\pm$ and (\textit{ii}) if $B < 0$, then $f_\pm < f_0$. Therefore the condition of existence of a topological vacuum imposes that the (possibly) physical vacuum should be restricted to the regions where $\C > 0$.

The requirement that $\s_0$  is physical vacuum, leads to some additional constraints on $D$ as well. Depending on the regions of $\la$ and $\m$ chosen we find that, for example, when  $0 < \la < 1/4$ and  for  the $\m_+$ model the following restriction 
$$0 < D^2 < \frac{(d-2)^2}{ \la L_{0+}^2} \left(1 - \sqrt{1 - 3\la}\right) ,$$ holds.
By changing the sign of $D$ and further considering the $\m_-$ model, we find that the condition that $D$ must obey is given by:
$$D^2 > \frac{(d-2)^2}{ \la \tilde{L}_{0+}^2} \left(1 - \sqrt{1 - 3\la}\right). $$

The solution for the equations of motion can be found following the same steps as we did in the case of quartic superpotential. We first take the ratio of eqs. (\ref{sys}) which gives:
$$\frac{d \s}{d A} = - \frac{4 B^4 L_+^2 L_-^2}{(d-2)^4} \; \frac{\s}{\s^2 + D/B} \; \prod_{i = 1}^4 ( \s^2 - \s^2_i ) \; , $$
where the $\s_i$ denote the algebraic solutions of (\ref{sigma top eq quad}). After integration, this yields
\begin{equation}
e^{A(\s)} = e^{A_\infty} |\s |^{- \frac{1}{s_0}} \; \prod_{i = 1}^{4}  \mid \! \s^2 - \s_i^2 \! \mid^{- 1/s_{top}^i} \; , \label{Solution quadratic}
\end{equation}
where the critical indexes (coinciding with(\ref{criti})) are now  given by
\begin{eqnarray}
&&s_0 = 4 B L_0 ( 1 - L_+^2 / L_0^2 ) ( 1 - L_-^2 / L_0^2 ) ; \;\;  s_{top}^p = - 16 (d-2)^{-1}  B^2 L_+^2 \s_p^2 ( 1 - L_-^2 / L_+^2 ) ; \label{s top + quad} \\
&&s_{top}^q = - 16 (d-2)^{-1}  B^2 L_-^2 \s_q^2 ( 1 - L_+^2 / L_-^2 ) ; \; p = 1,2 \; ; \; \; q = 3,4 \; .\label{s top - quad} 
\end{eqnarray}
As in the case of the quartic  superpotential, one can show that again the ``magic resonance" property   $s_0^{-1} + 2 \sum_i (s_{top}^i)^{-1} = 0$ (that determines the finite  $A(\infty) = A_\infty$ asymptotic of the scale factor) takes place, i.e.  there exist singular Lovelock DWs ending at one naked singularity .

In the region $\m > 0$ of parameter space (where only $L_+$ is real) the  singularities of the scale factor  (\ref{Solution quadratic}) are indeed placed at $\s_0$ and $\s_p$, $p = 1,2$. When $\m < 0$ and $\la > 0$ both topological scales $L_\pm$ are real, and then all the $\s_i$, $i= 1, \dots, 4$ represents ``topological vacua" singularities. Finally  for $\la$, $\m < 0$ both scales $L_\pm$ are complex and the unique scale factor singularity is at $\s_0$. 

The possible types of DWs chains for the quadratic superpotential are  more limited than for the quartic superpotential (largely discussed in Sect.4). Consider first the regions of parameter space when there is at least one topological vacuum. In this case, for GB-like physical regions (i.e. $B > 0$), due to the fact  there is only one physical vacuum, the Lovelock DWs scale factor represents two type of DWs of common boundaries: (\emph{i}) of "phys-top" AdS$_d$($\s_0$)/AdS$_d$($\s_{top}$) kind  or (\emph{ii}) a particular  M$_d$($\s_{0}$)/AdS$_d$($\s_{top}$) ``chain'' when  $D = 0$ (as we have explained in Sect3.2 we are not interested in). In the maximal scale region (i.e. $B < 0$) we have only one set of consistent b.c.'s representing two DWs  forming again one AdS$_d$($\s_{0}$)/AdS$_d$($\s_{top}$) ``chain''. In the regions of parameter space when there is no topological vacuum (i.e. for $\m$ and $\la$  both negatives), the only possibility represent a ``singular" DW solution, interpolating between the physical vacuum $\s_0$ and a naked singularity.

When the two topological vacua coincide, i.e. when the curve $h(f)$ has an inflection point, the gravitational theory turns out to become equivalent to the (dimensionally extended) AdS- Chern-Simons gravity (see for example \cite{zanelli},\cite{espanha}).
This degeneration corresponds to  the limit when $L_+ = L_- \equiv L_{CS}$, and therefore we can describe it by a procedure analogous to the one in Sect.4. Particularly, it can be seen from (\ref{Solution quadratic}), (\ref{s top + quad}) and (\ref{s top - quad}) that taking $s_p , s_q \rightarrow 0$ with $s_p / s_q < 0$, one finds
\begin{equation}
e^{A(\s)} \sim \prod_{i = 1}^2 \exp \left\{  \zeta_i \,  \frac{1}{\s^2 - \s_i^2} \right\} \; ; \;\; \zeta_i =  \frac{ (d-2)^2}{32 B^3 L_{CS}^3 \s_i^2} .
\end{equation}

Due to the fact that such Lovelock DWs solution is  ``marginally" degenerated (i.e.  critical exponent $s = 0$ corresponds to a marginal operator in the corresponding dual $CFT_{d-1}$) again as in the case of the quartic superpotential the interpretation  the vacuum as a boundary or a horizon depends on whether $\s$ approaches $\s_i$ from the right or the left, being a border if approaching from the right and a horizon if approaching from the left, i.e. preserving the nature of the vacuum before the limit $\e \rightarrow 0$ has been  taken.


\end{document}